\documentclass[sigconf]{acmart}

\usepackage{algpseudocode}
\usepackage{algorithm}
\algnewcommand\algorithmicforeach{\textbf{for each}}
\algdef{S}[FOR]{ForEach}[1]{\algorithmicforeach\ #1\ \algorithmicdo}

\usepackage{amsmath,tabularx}
\usepackage{amsthm}
\makeatletter
\newtheorem*{rep@theorem}{\rep@title}
\newcommand{\newreptheorem}[2]{%
\newenvironment{rep#1}[1]{%
 \def\rep@title{#2 \ref{##1}}%
 \begin{rep@theorem}}%
 {\end{rep@theorem}}}
\makeatother

\usepackage[labelformat=simple]{subcaption}

\usepackage[most]{tcolorbox}
\usepackage{multicol}
\usepackage{hhline}
\usepackage{tensor}
\usepackage{enumitem}
\usepackage{soul}

\newtheorem{theorem1}{Theorem}
\newreptheorem{theorem1}{Theorem}
\newtheorem{lemma1}{Lemma}
\newreptheorem{lemma1}{Lemma}

\newreptheorem{corollary1}{Corollary}
\newtheorem{proposition1}{Proposition}
\newreptheorem{proposition1}{Proposition}
\newtheorem{definition1}{Definition}

\usepackage{colortbl}
\usepackage{multirow}
\usepackage{tikz}
\usepackage{pgfplots}
\usepackage{rotating}   
\usetikzlibrary{decorations.pathreplacing,arrows,automata,calc,shapes.geometric}

\pgfdeclaredecoration{lightning bolt}{draw}{
\state{draw}[width=\pgfdecoratedpathlength]{
  \pgfpathmoveto{\pgfpointorigin}%
  \pgfpathlineto{\pgfpoint{\pgfdecoratedpathlength*0.6}%
    {-\pgfdecoratedpathlength*.1}}%
  \pgfpathlineto{\pgfpoint{\pgfdecoratedpathlength*0.55}{0pt}}%
  \pgfpathlineto{\pgfpoint{\pgfdecoratedpathlength}{0pt}}%
  \pgfpathlineto{\pgfpoint{\pgfdecoratedpathlength*0.4}%
    {\pgfdecoratedpathlength*.1}}%
  \pgfpathlineto{\pgfpoint{\pgfdecoratedpathlength*0.45}{0pt}}%
  \pgfpathclose%
}%
}

\usepackage{bm}

\makeatletter
\newcommand{\multiline}[1]{%
  \begin{tabularx}{\dimexpr\linewidth-\ALG@thistlm}[t]{@{}X@{}}
    #1
  \end{tabularx}
}
\makeatother

\makeatletter
\def\SOUL@hlpreamble{%
    \setul{\dp\strutbox}{\dimexpr\ht\strutbox+\dp\strutbox\relax}%
    \let\SOUL@stcolor\SOUL@hlcolor
    \SOUL@stpreamble
}
\makeatother

\let\oldnl\nl
\newcommand{\nonl}{\renewcommand{\nl}{\let\nl\oldnl}}

\acmConference[]{}{}{}

\begin{document}

\title[]{Root Causal Inference from Single Cell RNA Sequencing\\ with the Negative Binomial}


\author{Eric V. Strobl}
\email{}
 \affiliation{%
   \institution{}
   \city{}
   \state{}
   \country{}
 }

\renewcommand{\shortauthors}{Eric V. Strobl}

\begin{abstract}
Accurately inferring the root causes of disease from sequencing data can improve the discovery of novel therapeutic targets. However, existing root causal inference algorithms require perfectly measured continuous random variables. Single cell RNA sequencing (scRNA-seq) datasets contain large numbers of cells but non-negative counts measured by an \textit{error prone} process. We therefore introduce an algorithm called Root Causal Inference with Negative Binomials (RCI-NB) that accounts for count-based measurement error by separating negative binomial distributions into their gamma and Poisson components; the gamma distributions form a fully identifiable but latent post non-linear causal model representing the true RNA expression levels, which we only observe with Poisson corruption. RCI-NB identifies patient-specific root causal contributions from scRNA-seq datasets by integrating novel sparse regression and goodness of fit testing procedures that bypass Poisson measurement error. Experiments demonstrate significant improvements over existing alternatives.
\end{abstract}


\maketitle

\section{Introduction}

Causal inference algorithms identify causal relations from data. Most investigators infer causation using randomized controlled trials (RCTs). However, an RCT cannot distinguish between a cause and a \textit{root cause} of disease, or the initial perturbation to a biological system that ultimately induces a diagnostic label. Identifying the root causes of disease is critical for (a) understanding disease mechanisms and (b) discovering drug targets that treat disease at its biological onset.

Single cell RNA sequencing (scRNA-seq) datasets represent prime targets for root causal inference because they provide global but fine-grained snapshots of gene expression with ample numbers of cells. scRNA-seq also provides a functional read-out more proximal to the clinical phenotype than single nucleotide polymorphisms. Accurately inferring patient-specific root causes from scRNA-seq therefore has the potential to improve the discovery of novel therapeutic targets that significantly impact patient symptoms.

Unfortunately, most existing root causal inference algorithms assume perfectly measured, continuous random variables \cite{Strobl22a,Strobl22b,Strobl23}. Sequencing datasets contain counts measured by an error-prone sequencing process. Moreover, modern single cell pipelines cannot replicate the same measurements per cell \cite{Arzalluz17}. Customized methods that appropriately account for non-negativity and measurement error -- without relying on technical replicates -- have the potential to substantially improve the performance of existing methods from scRNA-seq.

The negative binomial distribution models counts as a mixture of the Poisson and gamma distributions, where the Poisson component can represent measurement error and the gamma distribution the expression level of an RNA molecule. The negative binomial also fits scRNA-seq data well by accounting for overdisperson and a high proportion of zeros \cite{Choudhary22,Svensson20}. As a result, many scientists analyze RNA-seq data with the negative binomial in the context of regression, normalization or differential hypothesis testing \cite{Hafemeister19,He21}. None however have utilized the negative binomial for identifying the root causes of disease. 
\begin{tcolorbox}[breakable,enhanced,frame hidden]
We therefore extend the negative binomial to root causal inference as follows:
\begin{enumerate}[leftmargin=*,label=(\arabic*)]
\item We propose a post-nonlinear causal model with gamma distributed error terms representing true continuous RNA expression levels. We can however only measure the expression levels as counts using a noisy sequencing process, which we model by the Poisson. The resultant Poisson-gamma mixture is the negative binomial.
\item We introduce a negative binomial regression procedure and goodness of fit hypothesis test that both bypass Poisson measurement error without technical replicates.
\item We integrate the regression procedure into an algorithm that identifies the parameters of the gamma distributions and latent causal graph.
\item We finally utilize the recovered parameters to identify the root causes of disease unique to each patient.
\end{enumerate}
\end{tcolorbox}
\noindent The resultant method called Root Causal Inference with Negative Binomials (RCI-NB) identifies patient-specific root causes of disease more accurately than existing alternatives from both simulated and real scRNA-seq datasets.

\section{Background}

\subsection{Structural Equations}

We can formalize causal inference under the framework of structural equation models (SEMs), or a set of deterministic equations over $p$ random variables $\widetilde{\bm{X}}$ such that:
\begin{equation} \label{eq:SEM}
\widetilde{X}_i = f_i(\textnormal{Pa}(\widetilde{X}_i),E_i), \hspace{5mm} \forall \widetilde{X}_i \in \widetilde{\bm{X}}.
\end{equation}
The random vector $\bm{E}$ denotes a set of mutually independent error terms, and $\textnormal{Pa}(\widetilde{X}_i) \subseteq \widetilde{\bm{X}} \setminus \widetilde{X}_i$ the parents, or direct causes, of $\widetilde{X}_i$. We can therefore associate a \textit{directed graph} $\mathbb{G}$ over $\widetilde{\bm{X}}$ to an SEM by drawing a directed edge from each member of $\textnormal{Pa}(\widetilde{X}_i)$ to $\widetilde{X}_i$. A \textit{directed path} in $\mathbb{G}$ from $\widetilde{X}_i$ to $\widetilde{X}_j$ refers to a sequence of adjacent directed edges from $\widetilde{X}_i$ to $\widetilde{X}_j$. We say that $\widetilde{X}_i$ is an \textit{ancestor} of $\widetilde{X}_j$ in $\mathbb{G}$ if there exists a directed path from $\widetilde{X}_i$ to $\widetilde{X}_j$ or $\widetilde{X}_i = \widetilde{X}_j$; similarly, $\widetilde{X}_j$ is a \textit{descendant} of $\widetilde{X}_i$. A \textit{cycle} occurs when $X_i$ is an ancestor of $\widetilde{X}_j$ and we have $\widetilde{X}_j \rightarrow \widetilde{X}_i$. We call $\mathbb{G}$ a \textit{directed acyclic graph} (DAG) if it contains no cycles. The joint distribution $\mathbb{P}_{\widetilde{\bm{X}}}$ over $\widetilde{\bm{X}}$ satisfies the \textit{causal Markov condition} if every variable in $\widetilde{\bm{X}}$ is independent of its non-descendants given its parents. Furthermore, $\mathbb{P}_{\widetilde{\bm{X}}}$ is \textit{causally minimal} if it satisfies the causal Markov condition relative to $\mathbb{G}$ but not to any proper sub-graph of $\mathbb{G}$.

\subsection{Related Work} \label{sec:RW}

Root causal analysis refers to a suite of methods designed to detect the root causes of undesired outcomes, typically in man-made systems within the industrial or healthcare industry \cite{Anderson06,Wu08}. The methods require a painstaking manual approach that implicitly or explicitly reconstructs the underlying causal graph. Strategies also rely on participants with deep knowledge of the underlying causal processes and therefore falter when applied to biological systems that remain largely unknown. 

A second line of work takes a similar approach by assuming a known set of structural equations but formalizes root causal analysis using the error terms of SEMs \cite{Budhathoki22,Budhathoki21}. These works unfortunately do not define patient-specific root causes of disease properly. For example, Root Causal Analysis of Outliers recovers root causal contribution scores for symptoms that are \textit{worse} than the symptoms of a given patient \cite{Budhathoki22}. We do not want to eliminate just the worse symptoms of a patient, but all of his symptoms. Attempting to correct the method with a predetermined cut-off score unfortunately foregoes patient-specificity. The Model Substitution
algorithm proposed in \cite{Budhathoki21} also loses specificity by identifying the root causes of changes in the \textit{marginal} distribution of the diagnosis. Moreover, both MS and RCAO assume that the user has knowledge of the structural equations and the ``normal'' counterfactual distributions of the error terms. The methods further require that the diagnosis correspond to a noiseless cutoff score, even though a diagnosis is noisy because it depends on the diagnostician in practice. RCAO and MS therefore utilize improper definitions of patient-specific root causes of disease and require a noiseless label, a known SEM as well as known counterfactual error term distributions.

A third line of work instead identifies patient-specific root causes of disease using the \textit{conditional} distribution of the diagnosis given the error terms. The authors do not require access to the underlying structural equations or error term distributions. \cite{Lasko19} performed independent component analysis (ICA) on electronic health record data and correctly recovered the top five root causes of hepatocellular carcinoma. The approach achieved clinical face validity, but the authors did not connect the strategy to causality. \cite{Strobl22a} later extended the idea to root causal analysis and introduced a more efficient algorithm called Root Causal Inference (RCI). The same authors later created a related procedure for handling latent confounding \cite{Strobl23}. All three of these algorithms assume linear relationships and continuous additive noise. Investigators thus later extended the work to the non-linear setting with the heteroscedastic noise model that allows non-linear conditional expectations and variances \cite{Strobl22b}. Unfortunately, even the non-linear approach assumes continuous random variables and no measurement error. The above algorithms therefore perform poorly when directly run on scRNA-seq datasets.

We improve on the aforementioned works by introducing an algorithm called Root Causal Inference with Negative Binomials (RCI-NB) that accounts for the measurement error and counts of scRNA-seq by bypassing the Poisson. The algorithm utilizes novel simulation-based regression and goodness of fit testing procedures. RCI-NB automatically recovers all parameters needed for the simulations using a top-down procedure introduced in Section \ref{sec:CD}. As a result, the algorithm requires no prior knowledge about the underlying structural equations or counterfactual distributions. Furthermore, RCI-NB allows a noisy label and maintains patient-specificity by identifying changes in the conditional distribution of the diagnosis.

\section{Negative Binomial Model}

We begin the development of RCI-NB by introducing a negative binomial SEM. We model the expression levels of RNA molecules $\widetilde{\bm{X}}$ using the following post non-linear SEM:
\begin{equation} \label{eq:PNL}
\widetilde{X}_i = \textnormal{exp}( \widetilde{\bm{X}} \beta_{\cdot i} ) E_i = \textnormal{exp}( \widetilde{\bm{X}} \beta_{\cdot i} + \textnormal{ln}(E_i)),
\end{equation}
for each $\widetilde{X}_i \in \widetilde{\bm{X}}$ similar to Equation \eqref{eq:SEM}, where \textit{post non-linearity} refers to the outer exponentiation. Exponentiation ensures that all variables in $\widetilde{\bm{X}}$ are positive and enforces faithfulness to the inverse canonical link function of the negative binomial generalized linear model \cite{Nelder72}. The entry $\beta_{ji} \not = 0$ if and only if $X_j \in \textnormal{Pa}(\widetilde{X}_i)$. We write $\beta_{\cdot i}$ to refer to the $i^\textnormal{th}$ column of $\beta$, and $\beta_{\widetilde{\bm{A}} i}$ to rows associated with $\widetilde{\bm{A}} \subseteq \widetilde{\bm{X}}$ in the $i^\textnormal{th}$ column. \cite{Zhang09} proved full identifiability of $\beta$ except in a few scenarios not applicable to this work.

Many RNA molecules have low expression levels. The gamma distribution places larger probability mass near zero than the log-normal with equal mean and variance. We therefore further assume that each $E_i \in \bm{E}$ follows the gamma distribution $\Gamma(r_i,r_i/\textnormal{exp}(\bm{P} \gamma_{\cdot i}))$ with shape $r_i$ and rate $r_i/\textnormal{exp}(\bm{P} \gamma_{\cdot i})$. The set $\bm{P}$ contains $q$ binary variables each indicating a patient from which we harvest cells. The error terms are therefore mutually independent given $\bm{P}$, or within each patient. 

We unfortunately cannot observe $\widetilde{X}$ in practice. Sequencing technologies instead approximate the expression level of each RNA by reverse transcribing and amplifying the molecules. Most technologies then count the number of complementary DNA sequences that align to a reference genome \cite{Hwang18}. As a result, sequencing technologies such as scRNA-seq only approximate reference RNA expression levels by counts \cite{Sarkar21}. 

The efficiency of the above process may differ between cells depending on e.g., cell diameter and the amount of reagents used. We can also only detect a small proportion of RNA molecules existing in a cell in general \cite{Sarkar21}. We therefore apply the law of rare events \cite{Papoulis02} and henceforth assume that we observe Poisson-corrupted counts $\bm{X}$ with each $X_i \in \bm{X}$ drawn according to:
\begin{equation} 
\nonumber
X_i \sim \textnormal{Pois}(\widetilde{X}_i C)
= \frac{(\widetilde{X}_i C)^{X_i}\textnormal{exp}(-\widetilde{X}_i C)}{X_i!}.
\end{equation}
The random variable $C>0$ denotes the cell-specific efficiencies of the sequencing process. The efficiencies differ due to the \textit{technology} -- not due to the biological system modeled by Equation \eqref{eq:PNL}. We can therefore approximate $C$ to high accuracy by a variety of control methodologies such as estimated library sizes or RNA spike-ins \cite{Ziegenhain22}.

Recall that $\widetilde{X}_i =\textnormal{exp}(\widetilde{\bm{X}}\beta_{\cdot i})E_i$ from Equation \eqref{eq:PNL}. We derive the conditional distribution of $X_i$ given $\textnormal{Pa}(\widetilde{X}_i) \cup \bm{P} \cup C$ by marginalizing out $\Gamma(r_i,r_i)$. The resultant Poisson-gamma mixture, or negative binomial distribution, obeys the probability mass function: 
\begin{equation} \label{eq:NB}
\mathbb{P}(X_i | \textnormal{Pa}(\widetilde{X}_i), \bm{P},C) =   \frac{\Gamma(X_i+r_i)}{\Gamma(X_i + 1) \Gamma(r_i)}\left( \frac{r_i}{r_i+\mu_i}\right)^{r_i}\left( \frac{\mu_i}{r_i+\mu_i} \right)^{X_i}.
\end{equation}
with dispersion parameter $r_i \in \bm{r}$, conditional expectation $\mu_i=\textnormal{exp}(\widetilde{\bm{X}}\beta_{\cdot i} +  \bm{P} \gamma_{\cdot i})C$ and variance $\mu_i + \frac{1}{r_i}\mu_i^2$. Several groups have shown that the quadratic variance accurately accounts for the overdispersion seen in real scRNA-seq data \cite{Choudhary22,Svensson20}. We will drop the subscripts of $r_i$ and $\mu_i$ to prevent notational cluttering, when it is clear that we focus on one $X_i \in \bm{X}$. 

In summary, we assume $\widetilde{\bm{X}}$ follows the fully identifiable SEM in Equation \eqref{eq:PNL} with gamma distributed error terms. We however can only observe $\widetilde{\bm{X}}$ with Poisson measurement error -- denoted by $\bm{X}$. We now seek to recover $(\beta,\bm{r},\gamma)$ from $\bm{X} \cup \bm{P} \cup C$ alone using negative binomial regression and goodness of fit testing, which we describe in the next two sections.

\section{Negative Binomial Regression} \label{sec:reg}
\subsection{Corrected Score Equations}
We first develop a negative binomial regression procedure that bypasses Poisson measurement error among the predictors. Most existing negative binomial regressors erroneously assume perfectly measured predictors or only Gaussian measurement error \cite{Guo01}. 

We in particular seek to regress $X_i$ on $\bm{P}$ and the perfectly measured non-descendants of $\widetilde{X}_i$, denoted by $\widetilde{\bm{A}} \subseteq \widetilde{\bm{X}} \setminus \widetilde{X}_i$, but only have access to the Poisson corrupted counterparts $\bm{A}$. Let $\bm{Z} = (\bm{A}/C, \bm{P})$ and $\widetilde{\bm{Z}} = (\widetilde{\bm{A}}, \bm{P})$. Further let $\alpha = (\beta_{\widetilde{\bm{A}} i}, \gamma_{\cdot i})^T$. We can then write the logarithm of the negative binomial probability mass function $\widetilde{L}(\alpha, r)$ as follows:
\begin{equation} \label{eq:logL}
\begin{aligned}
\textnormal{ln} \frac{\Gamma(X_i + r)}{\Gamma(X_i + 1) \Gamma(r)} + r \textnormal{ln}(r) + X_i \widetilde{\bm{Z}} \alpha - (X_i + r) \textnormal{ln}(r+\mu).
\end{aligned}
\end{equation}
Directly maximizing the expectation of the above expression requires access to $\widetilde{\bm{Z}}$. \cite{Nakamura90} showed that, if we can construct a \textit{corrected} function $L(\alpha,r)$ where:
 \begin{equation} \label{eq:corrected}
\mathbb{E}[L(\alpha,r)|X_i, \widetilde{\bm{Z}},C] = \widetilde{L}(\alpha,r),
 \end{equation}
then maximizing the (unconditional) expectation of $L(\alpha,r)$ still yields unbiased estimates of $\alpha$ and $r$. Observe that $\mathbb{E}(X_i\bm{Z} | X_i, \widetilde{\bm{Z}},C) = X_i\widetilde{\bm{Z}} $ for the term $X_i \widetilde{\bm{Z}} \alpha$ in Expression \eqref{eq:logL}, so $L(\alpha,r)$ satisfies:
\begin{equation} \nonumber
\begin{aligned}
\textnormal{ln} \frac{\Gamma(X_i + r)}{\Gamma(X_i + 1) \Gamma(r)} +r \textnormal{ln}(r) + X_i\bm{Z} \alpha - f(X_i, \bm{Z}, C),
\end{aligned}
\end{equation}
such that $\mathbb{E}(f|X_i, \widetilde{\bm{Z}},C) = (X_i + r) \textnormal{ln}(r+\mu)$ for some function $f$.

We find $f$ difficult to derive analytically. We can however simplify $(X_i + r) \textnormal{ln}(r+\mu)$ in Expression \eqref{eq:logL} as follows:
\begin{equation} \nonumber
\begin{aligned}
&\mathbb{E}_{X_i \widetilde{\bm{Z}} C} \left((X_i + r) \textnormal{ln}(r+\mu) \right) =\hspace{1mm} \mathbb{E}_{\widetilde{\bm{Z}}C} 
\left(\mathbb{E}_{X_i|\widetilde{\bm{Z}}C} (X_i + r) \textnormal{ln}(r+\mu)\right)\\
=\hspace{1mm}&\mathbb{E}_{\widetilde{\bm{Z}}C}\left((\mu + r)\textnormal{ln}(r+\mu)\right), 
\end{aligned}
\end{equation}
so that we can approximate the last expectation by averaging over $s$ samples drawn from the density $p(\widetilde{\bm{Z}},C)=p(\widetilde{\bm{Z}})p(C)$. We will show how to estimate $p(\widetilde{\bm{Z}},C)$ from data in Section \ref{sec:CD}. We therefore equivalently consider the following corrected function $L(\alpha,r)$:
\begin{equation} \label{eq:NB_sim1}
\textnormal{ln} \frac{\Gamma(X_i + r)}{\Gamma(X_i + 1) \Gamma(r)} +r \textnormal{ln}(r) + X_i\bm{Z} \alpha - \mathbb{E}\left((\mu + r)\textnormal{ln}(r+\mu)\right),
\end{equation}
which satisfies Equation \eqref{eq:corrected} as required.

We then set the expectation of the derivatives of $L(\alpha,r)$ to zero:
\begin{equation} \label{eq:asymp}
\begin{aligned}
\alpha &: \mathbb{E}\left(X_i \bm{Z}\right) - \mathbb{E}\left(\mu \widetilde{\bm{Z}}\right) = 0,\\
r &: \mathbb{E}\psi(X_i + r) - \psi(r) + \textnormal{ln}(r) - \mathbb{E} \textnormal{ln}(r + \mu) = 0,
\end{aligned}
\end{equation}
where $\psi$ denotes the digamma function. We replace the expectations with sample averages and quickly obtain the roots $(\alpha_n,r_n) = \theta_n$ of the corresponding score equations with $n$ samples by the Newton-Raphson method. Let $\theta_0$ denote the ground truth parameter values. The proposed approach achieves asymptotic normality:
\begin{proposition1} \label{prop:normality}
(Asymptotic normality) Assume $n \rightarrow \infty, s \rightarrow \infty$ and $n/s \rightarrow 0$. Further assume that $\textnormal{Var}(\mu \widetilde{\bm{Z}}, \textnormal{ln}(r+\mu))$ and $\Sigma = -\mathbb{E} S^\prime(\theta_0)$ are positive definite. Then $\sqrt{n}(\theta_n - \theta_0) \rightarrow \mathcal{N}(0,\Sigma^{-1}(J_1 + J_2 + J_3) \Sigma^{-1}).$
\end{proposition1}
\noindent We define $S^\prime, J_1, J_2$ and $J_3$ as well as detail longer proofs in the Appendix.

\subsection{Regularization} \label{sec:sparse}

Causal graphs in biology are frequently sparse, so we next introduce sparsity promoting regularization into the above negative binomial regressor. The equation for $\alpha$ in Equation \eqref{eq:asymp} does not depend on $r$ and is asymptotically equivalent to the score equation of the negative binomial with $r$ fixed. Recall that the negative binomial is a member of the exponential family with $r$ fixed. We therefore introduce regularization via the Bayesian information criterion (BIC) score \cite{Schwarz78}. 

Let $L_j(\alpha,r)$ denote the corrected log-likelihood for sample $j$. We maximize:
\begin{equation} \label{eq:BIC}
\left(\frac{1}{n} \sum_{j=1}^n L_j(\alpha,r)\right) - \frac{\lambda_n}{2} (\| \beta_{\widetilde{\bm{A}} i}\|_0 + \| \gamma_{\cdot i} - \widebar{\gamma}_{\cdot i}\|_2^2),
\end{equation}
where $\lambda_n = \textnormal{ln}(n)/n$ according to BIC and $\widebar{\gamma}_{\cdot i} = \frac{1}{q}\sum_{k=1}^q \gamma_{ki}$. We optimize the above expression quickly by customizing the expectation-maximization (EM) approach proposed in \cite{Liu16,Liu17}. The following equivalence relation holds:
\begin{equation} \nonumber
\|\beta_{\widetilde{\bm{A}} i}\|_0 = \sum_{j \in R} \frac{\beta_{j i}^2 }{ \beta_{j i}^2 }= \sum_{j \in R} \frac{\beta_{j i}^2}{ \eta_{j}^2} = \left\| \frac{\beta_{Ri}}{\eta_R} \right\|^2_2,
\end{equation}
where $\eta = (|\beta_{\widetilde{\bm{A}} i}|,\bm{1})$ and $R$ indexes the non-zero elements in $\beta_{\widetilde{\bm{A}} i}$. We collect $\beta_{R i}$ into the first $|R|$ entries of $\beta_{\widetilde{\bm{A}} i}$ for ease of notation. The ones in $\eta$ correspond to $\gamma_{\cdot i}$. Assume now that $\eta$ is a latent variable. The EM algorithm successively approximates $\alpha$ by iterating between expectation:
\begin{equation} \nonumber
\left(\frac{1}{n} \sum_{j=1}^n L_j(\alpha,r)\right) - \frac{\lambda_n}{2} \left( \left\| \frac{\beta_{Ri}}{\eta_R} \right\|^2_2 + \| \gamma_{\cdot i} - \widebar{\gamma}_{\cdot i} \|_2^2 \right),
\end{equation}
and maximization via the equation:
\begin{equation} \nonumber
\frac{1}{n} \sum_{j=1}^n x_{ij}\bm{z}_j  - \frac{1}{s} \sum_{j=1}^s  \mu_j\widetilde{\bm{z}}_j   - \lambda_n \left( \frac{\beta_{Ri}}{\eta_R}, 0,\gamma_{\cdot i} - \widebar{\gamma}_{\cdot i}\right) = 0.
\end{equation}
The zero vector on the left hand side corresponds to elements not in $R$. The above equation is potentially unstable due to division by entries in $\eta_R$ close to zero. Further, we do not know the indices $R$ in practice. We resolve both of these issues by element-wise multiplying both sides of the score equation by $\eta$ and instead solve:
\begin{equation} \label{score_EM}
\left(\frac{1}{n} \sum_{j=1}^n  x_{ij}\bm{z}_j  - \frac{1}{s} \sum_{j=1}^s \mu_j\widetilde{\bm{z}}_j  \right) \odot \eta  - \lambda_n \left(\beta_{\widetilde{\bm{A}} i}, \gamma_{\cdot i} - \widebar{\gamma}_{\cdot i}\right) = 0.
\end{equation}
We summarize the EM algorithm in Algorithm \ref{alg_DL}; it almost always converges with a finite number of samples in practice.

\begin{algorithm}[]
 \hspace{-30.5mm}\textbf{Input:} $\varepsilon$ and samples of $\bm{Z}, \widetilde{\bm{Z}}, X_i, C$\\
 \hspace{-59mm}\textbf{Output:}  $\alpha, r$
\begin{algorithmic}[1]
\State $\alpha \leftarrow$ solve Equation \eqref{score_EM} with $\eta = (|\beta_{\widetilde{\bm{A}} i}|,\bm{1})$ and $|\beta_{\widetilde{\bm{A}} i}| = 1$
\Repeat
    \State $\alpha_0 \leftarrow$ solve Equation \eqref{score_EM} with $\eta = (|\beta_{\widetilde{\bm{A}} i}|,\bm{1})$
    \State $\Delta = \| \alpha_0 - \alpha\|_1$ \label{alg_DL:partial}
    \State $\alpha \leftarrow \alpha_0$
\Until{$\Delta < \varepsilon$}
\State $r \leftarrow$ solve for $r$ in Equation \eqref{eq:asymp} with $\alpha$ \label{alg_DL:r}
\end{algorithmic}
\caption{Neg. Binomial Expectation Maximization (NB-EM)} \label{alg_DL}
\end{algorithm}

\section{Goodness of Fit Test}

We have thus far assumed that $p(X_i|\widetilde{\bm{Z}},C)$ indeed follows a negative binomial distribution. We now address the problem of determining whether the negative binomial distribution holds in this section by constructing a score-based goodness of fit test.

Assume for now that we have access to $\widetilde{\bm{Z}}\cup C$ in order to compute $\mu$. We construct a hypothesis test with a flexible order-$k$ alternative probability mass function:
\begin{equation} \nonumber
\begin{aligned}
p_k(X_i|\widetilde{\bm{Z}},C) = N(h,\phi,\mu, r) \textnormal{exp}\left( \sum_{j=1}^k h_j(X_i,\mu, r) \phi_j \right) p_0(X_i|\widetilde{\bm{Z}},C),
\end{aligned}
\end{equation}
where $p_0(X_i|\widetilde{\bm{Z}},C)$ denotes the negative binomial probability mass function under the null hypothesis. We now suppress the inputs to some functions for cleaner exposition. The function $\textnormal{exp}( \sum_{j=1}^k h_j \phi_j )$ $=\textnormal{exp}(h\phi)$ is non-negative and equal to one under the null hypothesis that $\phi = 0$, or when the negative binomial $p_0(X_i|\widetilde{\bm{Z}},C)$ holds. The normalizing function $N$ ensures that $p_k$ integrates to one. Each function $h_j$ must have zero expectation under the null hypothesis, denoted by $\mathbb{E}_0(h_j)=0$. We will show how to intelligently choose such functions shortly.

We now take the expectation of the logarithm of $p_k$. The normalizing function $N$ has derivative $ \frac{\partial\textnormal{log} N}{\partial \phi}$ equal to $-\mathbb{E}_k(h)$, or the negative expectation under the order-$k$ alternative (Lemma 4.2.1 in \cite{Rayner09}). As a result, $ \frac{\partial\textnormal{log} N}{\partial \phi} = 0$ under the null hypothesis, so we can write the population score equation with respect to $\phi$ under the null as $\mathbb{E}_0(h) = 0.$ This implies that:
 \begin{equation} \label{eq:chi2}
U = \frac{1}{n} \sum_{j=1}^n h_j^T \widehat{\Pi}^{-1} h_j \leadsto \chi^2_k,
\end{equation}
by the central limit theorem. Here, we index the samples of $h$ rather than its entries. The matrix $\widehat{\Pi}$ denotes the sample covariance matrix of the vector $h$, which we will describe soon.

The $\chi^2$ test loses power with too many functions in $h$ and may not converge to the asymptotic distribution fast enough with large variances for realistic sample sizes. We therefore fix $k=2$ and utilize the bounded functions $h_j = m_j - \mathbb{E}(m_j | \mu, r)$, where $m_1 = \textnormal{exp}(-X_i) \in (0,1]$ and $m_2 = \textnormal{sin}(X_i) \in [-1,1]$. The conditional expectations of $m_1$ and $m_2$ admit closed forms under the negative binomial:
\begin{equation} \nonumber
\begin{aligned}
\mathbb{E}(m_1 | \mu, r) &= \textnormal{exp}(r) (r/((\textnormal{exp}(1) - 1) \mu + \textnormal{exp}(1) r))^r, \\
\mathbb{E}(m_2 | \mu, r) &= \frac{ir^r}{2} \left((-\textnormal{exp}(-i)\mu + \mu + r)^{-r} - (-\textnormal{exp}(i)\mu + \mu + r)^{-r}\right),
\end{aligned}
\end{equation}
where $i = \sqrt{-1}$.

Recall that we estimate $(\alpha, r) = \theta$ in practice by NB-EM. Boos \cite{Boos92} used a first-order Taylorian expansion to account for the non-maximum likelihood estimation using an adjusted covariance matrix $\Pi$ equal to:
\begin{equation} \nonumber
\begin{aligned}
B_{\phi\phi} - &A_{\phi\theta}A_{\theta \theta}^{-1}B_{\theta \phi} - B_{\phi \theta}(A_{\theta \theta}^{-1})^TA_{\phi \theta}^T + A_{\phi \theta}A_{\theta \theta}^{-1}B_{\theta \theta}(A_{\theta \theta}^{-1})^TA_{\phi \theta}^T
\end{aligned}
\end{equation}
where:
\begin{equation} \nonumber
\begin{aligned}
S = &\hspace{1mm} \left( h_{1}-\mathbb{E}(m_1 | \mu, r), h_{2}-\mathbb{E}(m_2 | \mu, r), X_i\bm{Z}-\mu\widetilde{\bm{Z}} \right)^T,\\
A = &\hspace{1mm}-\mathbb{E} \left( \frac{\partial S}{\partial (\phi,\theta)} \right),\hspace{5mm} B = \mathbb{E} \left(S S^T \right). 
\end{aligned}
\end{equation}

We must finally account for the fact that we observe $\bm{Z}$ but simulate $\widetilde{\bm{Z}}$. We split the functions in $S$ into two groups:
\begin{equation} \nonumber
\begin{aligned}
S_{\bm{Z}} = \left( h_{1}, h_{2}, X_i\bm{Z}  \right)^T, \hspace{0.25cm}
S_{\widetilde{\bm{Z}}} = \left( -\mathbb{E}(m_1 | \mu, r), -\mathbb{E}(m_2 | \mu, r), -\mu\widetilde{\bm{Z}} \right)^T,
\end{aligned}
\end{equation}
yielding the new matrices:
\begin{equation} \nonumber
\begin{aligned}
A = -\mathbb{E}_{\widetilde{\bm{Z}}} \left( \frac{\partial S_{\widetilde{\bm{Z}}}}{\partial (\phi,\theta)} \right), \hspace{0.25cm} B = \mathbb{E}_{\bm{Z}} \left(S_{\bm{Z}}S^T_{\bm{Z}} \right) +\mathbb{E}_{\widetilde{\bm{Z}}} \left(S_{\widetilde{\bm{Z}}}S_{\widetilde{\bm{Z}}}^T \right). 
\end{aligned}
\end{equation}
We then reject the null hypothesis that the negative binomial holds when the statistic $U$ in Equation \eqref{eq:chi2} falls above the critical value determined by the Type I error rate. 

\section{Causal Inference} \label{sec:CD}
\subsection{Parameter Estimation}
We have thus far created a sparse negative binomial regressor and score-based goodness of fit test that both bypass Poisson measurement error. They however require access to $p(\widetilde{\bm{Z}},C)$. We now design an algorithm called Recover Parameters (RP) that utilizes regression and goodness of fit testing to systematically identify the $\beta$, $\bm{r}$ and $\gamma$ parameters of $p(\widetilde{\bm{Z}},C)$. We summarize RP in Algorithm \ref{alg_CDGoF}. 

\begin{algorithm}[]
\hspace{-60mm} \textbf{Input:} $\bm{X}, \bm{P}, C$\\
\hspace{-58.5mm} \textbf{Output:}  $\beta, \bm{r}, \gamma$
\begin{algorithmic}[1]
\State $\bm{A} \leftarrow \emptyset$; $\beta \leftarrow 0_{p \times p}$; $\gamma \leftarrow 0_{q\times p}$; $\bm{r} \leftarrow 0_{p\times 1}$
\Repeat
    \ForEach{$X_i \in \bm{X}$}
        \State $\widetilde{\bm{A}} \leftarrow$ Simulate from $p(\widetilde{\bm{A}})$ with $(\beta_{\cdot \bm{A}},\bm{r}_{\bm{A}},\gamma_{\cdot \bm{A}})$ \label{alg_CDGoF:sim}
        \State $(\beta_{\bm{A}i},r_i,\gamma_{\cdot i}) \leftarrow$ Regress $X_i$ on $\widetilde{\bm{A}} \cup \bm{P}\cup C$ using NB-EM \label{alg_CDGoF:reg1}
        \If{ $X_i \sim$ negative binomial given $\widetilde{\bm{A}} \cup \bm{P} \cup C$ \label{alg_CDGoF:test1}}
            \State $\bm{A} \leftarrow \bm{A} \cup X_i$; $\bm{X} \leftarrow \bm{X} \setminus X_i$; \textbf{break} \label{alg_CDGoF:change}
        \EndIf
    \EndFor
\Until{$\bm{X} = \emptyset$}
\end{algorithmic}
\caption{Recover Parameters (RP)} \label{alg_CDGoF}
\end{algorithm}

RP performs causal discovery in a top-down fashion; the algorithm discovers the roots, then the children of the roots, and so forth. The algorithm first fits negative binomial distributions on each random variable given $\bm{P} \cup C$ in Line \ref{alg_CDGoF:reg1}. RP then tests whether each variable follows a negative binomial in Line \ref{alg_CDGoF:test1}. If a variable does, then RP places it into $\bm{A}$ and eliminates it from $\bm{X}$ in Line \ref{alg_CDGoF:change}. We eliminate the variable from $\bm{X}$ with the smallest $U$ statistic in practice to avoid dependence on a pre-specified Type I error rate.

When the negative binomial holds, the set $(\beta_{\widetilde{\bm{A}} i},r_i,\gamma_{\cdot i})$ obtained in Line \ref{alg_CDGoF:reg1} contains the gamma distribution parameters of $E_i$ because $E_i \sim \Gamma(r_i,r_i/\textnormal{exp}(\bm{P}\gamma_{\cdot i}))$. RP can therefore simulate values from $p(\widetilde{\bm{A}})$ in Line \ref{alg_CDGoF:sim} by drawing $\bm{P}$ as well as the samples of the corresponding error terms of $\widetilde{\bm{A}}$, and then passing the values through the structural equations associated with $\widetilde{\bm{A}}$ in Equation \eqref{eq:PNL}.

RP can now discover the children of the roots by independently sampling $C$ by bootstrap, regressing on $\widetilde{\bm{A}} \cup \bm{P} \cup C$ and testing whether the model fits a negative binomial. The algorithm repeats the above process of simulation, sparse regression and goodness of fit testing until it moves all variables from $\bm{X}$ into $\bm{A}$. The algorithm is formally sound and complete:
\begin{theorem1} \label{thm:identifiability}
(Identifiability) If $\mathbb{P}_{\widetilde{\bm{X}}|\bm{P}}$ is causally minimal and $X_i \sim \textnormal{Pois}(\widetilde{X}_i C)$ for each $X_i \in \bm{X}$, then RP recovers $(\beta,\bm{r},\gamma)$ with regression and goodness of fit oracles.
\end{theorem1}
\noindent We cannot reach the conclusion directly from the results of \cite{Zhang09} due to the Poisson measurement error. We therefore instead prove the theorem in the Appendix using an overdispersion score developed for quadratic variance functions \cite{Park17}.

 \subsection{Root Causal Contributions}

 \begin{figure*}[]
\centering
\begin{subfigure}{0.49\textwidth}
\centering
\raisebox{1.0cm}{\begin{tikzpicture}[scale=1.0, shorten >=1pt,auto,node distance=2.8cm, semithick,
  inj/.pic = {\draw (0,0) -- ++ (0,2mm) 
                node[minimum size=2mm, fill=red!60,above] {}
                node[draw, semithick, minimum width=2mm, minimum height=5mm,above] (aux) {};
              \draw[thick] (aux.west) -- (aux.east); 
              \draw[thick,{Bar[width=2mm]}-{Hooks[width=4mm]}] (aux.center) -- ++ (0,4mm) coordinate (-inj);
              }]
                    
\tikzset{vertex/.style = {inner sep=0.4pt}}
\tikzset{edge/.style = {->,> = latex'}}
 
\node[vertex] (1) at  (0,0) {$\widetilde{X}_1$};
\node[vertex] (2) at  (1.25,0) {$\widetilde{X}_2$};
\node[vertex] (3) at  (2.5,0.5) {$\widetilde{X}_3$};
\node[vertex] (4) at  (2.5,-0.5) {$\widetilde{X}_4$};
\node[vertex] (5) at  (3.75,0.5) {$\widetilde{X}_5$};
\node[vertex] (6) at  (3.75,-0.5) {$\widetilde{X}_6$};
\node[vertex] (7) at  (5,0) {$D$};

\fill [blue, decoration=lightning bolt, decorate] (1.25,0.25) -- ++ (0.55,0.65);

\draw[edge] (1) to (2);
\draw[edge,blue] (2) to (3);
\draw[edge,blue] (2) to (4);
\draw[edge,blue] (3) to (5);
\draw[edge,blue] (4) to (6);
\draw[edge,blue] (5) to (7);
\draw[edge,blue] (6) to (7);
\end{tikzpicture}}
\caption{} \label{fig_shock}
\end{subfigure} \begin{subfigure}{0.49\textwidth}
\centering
\begin{tikzpicture}[scale=1.0, shorten >=1pt,auto,node distance=2.8cm, semithick]
                    
\tikzset{vertex/.style = {inner sep=0.4pt}}
\tikzset{edge/.style = {->,> = latex'}}
 
\node[vertex] (1) at  (0,0) {$\widetilde{X}_1$};
\node[vertex,draw=blue, circle] (2) at  (1.25,0) {$\widetilde{X}_2$};
\node[vertex,draw=blue, circle] (3) at  (2.5,0.5) {$\widetilde{X}_3$};
\node[vertex,draw=blue, circle] (4) at  (2.5,-0.5) {$\widetilde{X}_4$};
\node[vertex,draw=blue, circle] (5) at  (3.75,0.5) {$\widetilde{X}_5$};
\node[vertex,draw=blue, circle] (6) at  (3.75,-0.5) {$\widetilde{X}_6$};
\node[vertex,draw=blue, circle, minimum size=0.50cm] (7) at  (5,0) {$D$};

\node[vertex] (8) at  (-0.5,1) {$E_1$};
\draw[edge] (8) to (1);
\node[vertex] (9) at  (0.75,1) {$E_2 = \textcolor{blue}{e_2}$};
\draw[edge,blue] (9) to (2);
\node[vertex] (10) at  (2,1.5) {$E_3$};
\draw[edge] (10) to (3);
\node[vertex] (11) at  (3.25,1.5) {$E_5$};
\draw[edge] (11) to (5);
\node[vertex] (13) at  (2,-1.5) {$E_4$};
\draw[edge] (13) to (4);
\node[vertex] (14) at  (3.25,-1.5) {$E_6$};
\draw[edge] (14) to (6);

\draw[edge] (1) to (2);
\draw[edge,blue] (2) to (3);
\draw[edge,blue] (2) to (4);
\draw[edge,blue] (3) to (5);
\draw[edge,blue] (4) to (6);
\draw[edge,blue] (5) to (7);
\draw[edge,blue] (6) to (7);
\end{tikzpicture}
\caption{} \label{fig_SEM_down}
\end{subfigure}
\caption{The lightning bolt in (a) corresponds to an exogenous perturbation of the random variable $\widetilde{X}_2$ that affects downstream variables ultimately inducing a diagnosis $D$ for a specific patient. In (b), we model the lightning bolt as an intervention on the error term $E_2$ to the value $e_2$; this in turn changes the value of all downstream variables including $D$.}
\end{figure*}
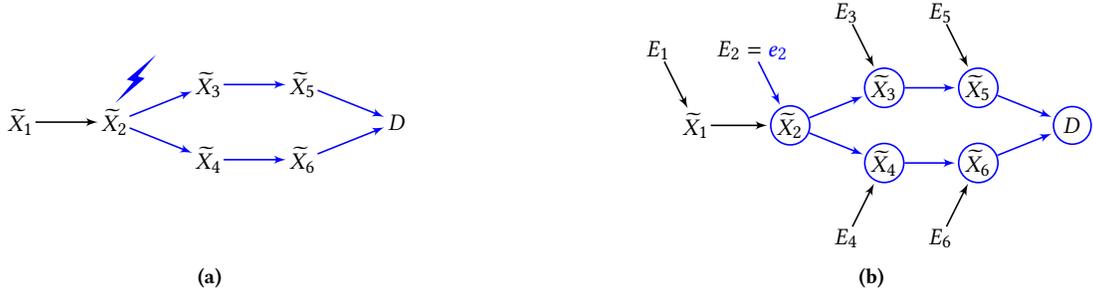

We would like to utilize the recovered parameters from RP to identify root causes of disease specific to each patient. A \textit{root cause of disease} intuitively corresponds to an initial perturbation to a biological system that ultimately induces a diagnostic label (Figure \ref{fig_shock}). We can formulate this intuition mathematically by first introducing a binary label $D$ labeling a sample with a diagnosis of a certain illness ($D=1$) or a healthy control ($D=0$). We assume that
$D$ is a terminal vertex such that $\mathbb{P}(D|\widetilde{\bm{X}},\bm{P}) = \textnormal{logistic}(\widetilde{\bm{X}} \beta_{\cdot D} + \bm{P} \gamma_{\cdot D})$.
The logistic function emphasizes that the diagnosis is a \textit{noisy} label of the predictors $\widetilde{\bm{X}}$. We may likewise consider other functions for a binary target, such as the probit. We can associate an error term $E_D$ to $D$ but reserve the notation $\bm{E}$ for the error terms of $\widetilde{\bm{X}}$ so that $E_D \not \in \bm{E}$.

A root cause of disease for a specific patient then corresponds to a natural intervention on the error term of an ancestor of $D$. In particular, consider $\widetilde{X}_i = \textnormal{exp}\left( \widetilde{\bm{X}} \beta_{\cdot i} \right) E_i$ from Equation \eqref{eq:NB} and suppose $E_i = \bar{e}_i$ for a healthy control.  We can interpret each error term $E_i \in \bm{E}$ as the combined effects of unobserved variables lying upstream of only $X_i$ -- such as the DNA sequence, acetylation or methylation status of a gene. An exogenous insult -- such as a somatic mutation or toxin -- then changes the value of $E_i$ from $\bar{e}_i$ to an ``unhealthy'' one $e_i$. The change of $E_i$ from $\bar{e}_i$ to $e_i$ affects downstream variables, ultimately impacting variables involved in the diagnostic criteria and therefore the diagnosis $D$ itself (Figure \ref{fig_SEM_down}).  

We can quantify the change in probability of $D$ using the following logarithmic odds:
\begin{equation} \label{eq:lin_LR}
D_0 = \textnormal{ln}\left( \frac{\mathbb{P}(D=1|\widetilde{\bm{X}},\bm{P})}{\mathbb{P}(D=0|\widetilde{\bm{X}},\bm{P})}\right) = \widetilde{\bm{X}}\beta_{\cdot D} + \bm{P} \gamma_{\cdot D}.
\end{equation}
The equation depends on the variables $\widetilde{\bm{X}} \cup \bm{P}$, but we would like to quantify the causal effect of each error term $E_i \in \bm{E}$ on $D$ for a specific patient.

The above logarithmic odds of the logistic regression model of $D$ on $\widetilde{\bm{X}} \cup \bm{P}$ admits a linear form. However, the logarithmic odds of $D$ on $\bm{E}$ for patient $j$, denoted by $f^j(\bm{E})$, generally requires a non-linear function. We learn $f^j(\bm{E})$ by performing non-linear logistic regression with the cells associated with patient $j$. Let $v^j(\bm{W})$ correspond to the conditional expectation of the non-linear model $\mathbb{E}(f^j(\bm{E})|\bm{W})$ for some $\bm{W} \subseteq \bm{E} \setminus E_i$. We can measure the change in probability when intervening on some $E_i \in \bm{E}$ for patient $j$ via the difference $\delta^j_{E_i \bm{W}} = v^j(E_i,\bm{W})-v^j(\bm{W})$. We have $\delta^j_{e_i \bm{w}} >0$ when $E_i = e_i$ increases the probability that $D=1$ because $v^j(e_i,\bm{w})$ is larger than $v^j(\bm{w})$. 

We do not a priori know which $\bm{W}$ to choose, so we average over all possible $\bm{W} \subseteq \bm{E} \setminus E_i$ as follows:
\begin{equation} \nonumber
    S^j_i = \frac{1}{p}\sum_{\bm{W} \subseteq (\bm{E} \setminus E_i)} \frac{1}{\binom{p-1} {|\bm{W}|}} \delta^j_{E_i \bm{W}}.
\end{equation}
An instantiation of the above quantity corresponds precisely to the Shapley value of \cite{Lundberg17} which, as the reader may recall, is the \textit{only} additive feature attribution measure satisfying the local accuracy, missingness and consistency desiderata. We can thus quantify the root causal contribution of $E_i$ on $D$ using $S^j_i$. 

Measurement error precludes recovery of the exact values of $\bm{E}$ and therefore of $S^j_i$. We instead compute the expected Shapley for patient $j$ given by $\mathbb{E}(S^j_i|D=1) = \Upsilon_i^j$. This expected Shapley also satisfies the three desiderata by linearity of expectation:
\begin{enumerate}[leftmargin=*,label=(\arabic*)]
    \item Local accuracy: $\sum_{i=1}^p \Upsilon_i^j = \mathbb{E} [f^j(\bm{E})|D=1] - \mathbb{E} f^j(\bm{E})$;
    \item Missingness: if $E_i \not \in \bm{E}$, then $\Upsilon_i^j = 0$;
    \item Consistency: We have $\ddot{\Upsilon}_i^j \geq \Upsilon_i^j$ for any two models $\ddot{f}^{j}$ and $f^j$ where $\mathbb{E}(\ddot{\delta}^j_{E_i\bm{W}}|D=1) \geq \mathbb{E}(\delta^j_{E_i\bm{W}}|D=1)$ for all $\bm{W} \subseteq \bm{E} \setminus E_i$.
\end{enumerate}
The first criterion ensures that the total score $\sum_{i=1}^p \Upsilon_i^j$ remains invariant to changes in the patient-specific disease prevalence rate $\mathbb{E} f^j(\bm{E})$. The second criterion implies $s_D = 0$ because $E_D \not \in \bm{E}$. The first and third criteria together imply that each $\Upsilon_{i}^j$ is also invariant to changes in the disease prevalence rate, since we must have $\ddot{\delta}^j_{E_i\bm{W}} \geq \delta^j_{E_i\bm{W}}$ for all $\bm{W} \subseteq \bm{E} \setminus E_i$ and $X_i \in \bm{X}$. The three desiderata are therefore necessary. 

We now introduce the following definition:
\begin{definition1}
The root causal contribution of $X_i$ for patient $j$ is $\Upsilon_i^j$. Similarly, $X_i$ is a \textit{root cause of disease} ($D=1$) for patient $j$ if $\Upsilon_i^j > 0$. 
\end{definition1}
\noindent $X_i$ is not a \textit{root cause of disease} for patient $j$ if $\Upsilon_i^j \leq 0$ because $E_i$ does not on average increase the probability that $D=1$ in this case.

\subsection{Root Causal Inference}
We estimate $\Upsilon_i^j$ for each variable $X_i \in \bm{X}$ and patient $P_j \in \bm{P}$ using the Root Causal Inference with Negative Binomials (RCI-NB) algorithm summarized in Algorithm \ref{alg:RCI_NB}. RCI-NB first runs RP in Line \ref{alg:RCI_NB:CD_GOF} to estimate the coefficients $\beta_{\cdot \widetilde{\bm{X}}}$ and gamma distribution parameters $(\bm{r},\gamma_{\cdot \widetilde{\bm{X}}})$ in order to simulate samples from   $p(\bm{E},\widetilde{\bm{X}},\bm{P})$. The algorithm then obtains $(\beta_{\cdot D}, \gamma_{\cdot D})$ by regressing $D$ on $\widetilde{\bm{X}} \cup \bm{P}$ with the Logistic Regression Expectation Maximization (LR-EM) algorithm in Line \ref{alg:RCI_NB:LR_EM}. LR-EM proceeds just like NB-EM but with Line \ref{alg_DL:r} removed and Equation \eqref{score_EM} replaced by:
\begin{equation} \nonumber
\left(\frac{1}{n} \sum_{j=1}^n d_j\bm{z}_j  - \frac{1}{s} \sum_{j=1}^s \frac{\mu_j \widetilde{\bm{z}}_j}{1+\mu_j} \right)\odot \eta - \lambda_n \left(\beta_{\cdot D}, \gamma_{\cdot D} - \widebar{\gamma}_{\cdot D}\right) = 0,
\end{equation}
or the corresponding score equations for logistic regression.
The recovered parameters in turn enable simulation of $D_0$. RCI-NB therefore non-linearly regresses $D_0$ on $\bm{E}$ for each patient $j$. We use XGBoost in this paper, so we can quickly compute the expected Shapley values for each patient using TreeSHAP in Line \ref{alg:RCI_NB:shap} \cite{Lundberg18}. 

\begin{algorithm}[]
 \hspace{-56mm} \textbf{Input:} $\bm{X},\bm{P},C,D$\\
 \hspace{-64.4mm}  \textbf{Output:}  $\Upsilon$
\begin{algorithmic}[1]
\State $(\beta_{\cdot \widetilde{\bm{X}}}, \bm{r}, \gamma_{\cdot \widetilde{\bm{X}}}) \leftarrow \textnormal{RP}(\bm{X},\bm{P},C)$ \label{alg:RCI_NB:CD_GOF}
\State $(\bm{E}, \widetilde{\bm{X}},\bm{P}) \leftarrow$ Simulate from  $p(\bm{E},\widetilde{\bm{X}},\bm{P})$  with $(\beta_{\cdot \widetilde{\bm{X}}}, \bm{r}, \gamma_{\cdot \widetilde{\bm{X}}})$ \label{alg:RCI_NB:sim}
\State $(\beta_{\cdot D}, \gamma_{\cdot D}) \leftarrow$ Regress $D$ on $\widetilde{\bm{X}} \cup \bm{P}$ using LR-EM \label{alg:RCI_NB:LR_EM}
\State $D_0 \leftarrow$ Simulate samples from Eq. \eqref{eq:lin_LR} using $(\widetilde{\bm{X}},\bm{P},\beta_{\cdot D}, \gamma_{\cdot D})$ \label{alg:RCI_NB:D0}
\State $\Upsilon_i^j \leftarrow \mathbb{E}(S_i^j|D=1)$ for each $X_i \in \bm{X}$ and patient $j$ \label{alg:RCI_NB:shap}
\end{algorithmic}
\caption{Root Causal Inference with Negative Binomials (RCI-NB)} \label{alg:RCI_NB}
\end{algorithm}

 \begin{table*}[]
\begin{subtable}{0.46\textwidth} 
\centering
\begin{tabular}{cc|ccccc}
\hhline{=======}
$p$                   & $n$      & RCI-NB         & RCI   & ICA  & ANM   & HNM  \\ \hline
\multirow{2}{*}{7} & 10,000  & \textbf{0.264}             & 0.331                   & 0.331                    & 0.342                   & 0.351                  \\
                    & 100,000 & \textbf{0.137}             & 0.329                   & 0.329                    & 0.338                   & 0.349                \\ \hline
\multirow{2}{*}{12} & 10,000  & \textbf{0.224}             & 0.265                   & 0.264                    & 0.269                   & 0.277                  \\
                    & 100,000 & \textbf{0.127}             & 0.266                   & 0.265                        & 0.269                       & 0.279    \\           
                    \hhline{=======}
\end{tabular}
\caption{} \label{exp:synth:RMSE}
\end{subtable}
\begin{subtable}{0.46\textwidth} 
\centering
\begin{tabular}{cc|ccccc}
\hhline{=======}
$p$                   & $n$      & RCI-NB         & RCI   & ICA  & ANM   & HNM  \\ \hline
\multirow{2}{*}{7} & 10,000  & 26.30             &  0.068                   & 0.061                    & 56.40                   & 44.30                  \\
                    & 100,000 & 328.4             & 0.867                   & 0.563                    & 579.0                   & 463.3                \\ \hline
\multirow{2}{*}{12} & 10,000  & 81.28             & 0.105                   & 0.103                    & 261.8                   & 231.4                  \\
                    & 100,000 & 1114             & 1.506                   & 1.021                        & 2701                       & 2365    \\           
                    \hhline{=======}
\end{tabular}
\caption{} \label{exp:synth:time}
\end{subtable}
\caption{Accuracy (a) and timing (b) results with the synthetic data. We multiplied the RMSE values by ten for ease of display. RCI-NB achieved the lowest mean RMSE in all cases and completed approximately twice as fast as ANM and HNM.} \label{exp:synth}
\end{table*}

A Shapley oracle outputs the true expected Shapley for each patient given $p(\bm{E},\bm{P},D_0)$. The RCI-NB algorithm is sound with oracle information:
\begin{theorem1}
RCI-NB outputs the true expected Shapley values, or $\Upsilon_i^j$ for each variable $X_i \in \bm{X}$ and patient $j$ given regression, goodness of fit and Shapley oracles.
\end{theorem1}
\begin{proof}
RP recovers the parameters $(\beta_{\cdot \widetilde{\bm{X}}}, \bm{r}, \gamma_{\cdot \widetilde{\bm{X}}})$ with negative binomial regression and goodness of fit oracles per Theorem \ref{thm:identifiability}. Similarly, RCI-NB recovers $\beta_{\cdot D}$ and $\gamma_{\cdot D}$ with a logistic regression oracle. Lines 2 and 4 therefore simulate samples from $p(\bm{E},\bm{P},D_0)$ and recover $\mathbb{E}(S_i^j | D=1)$ with a Shapley oracle in Line \ref{alg:RCI_NB:shap}.
\end{proof}

\section{Experiments}

\subsection{Algorithms}
We compared RCI-NB against the following four algorithms representing the state of the art in inference for patient-specific root causes of disease:
\begin{enumerate}[leftmargin=*,label=(\arabic*)]
    \item Root Causal Inference (RCI): an efficient top-down algorithm that infers patient-specific root causes assuming a linear model with non-Gaussian error terms \cite{Strobl22a}.
    \item Independent Component Analysis (ICA): utilizes a general purpose ICA algorithm to extract the error term values also assuming a linear model with non-Gaussian error terms \cite{Lasko19}.
    \item Generalized Root Causal Inference with the Additive Noise Model (ANM): a bottom-up algorithm that generalizes RCI to non-linear models \cite{Strobl22b}. We equipped GRCI with ANM by solving for $(\beta, \gamma)$ with NB-EM. We then subtracted out the conditional means to recover the error term values.
    \item Generalized Root Causal Inference with the Heteroscedastic Noise Model (HNM): same as ANM, but we solved for both $(\beta, \bm{r},\gamma)$ with NB-EM. We then subtracted out the conditional means and dividing by the conditional standard deviations to recover the error term values.
\end{enumerate}
We equipped RCI-NB, ANM and HNM with the same XGBoost TreeSHAP procedure for estimating the expected Shapley values with extracted error terms \cite{Lundberg18}. ANM and HNM both utilize NB-EM and LR-EM like RCI-NB. RCI and ICA use linear logistic regression models for inferring the expected Shapley values in accordance with their linearity assumption. No algorithm except RCI-NB takes measurement error into account.

\textbf{Reproducibility.} All code needed to replicate the experimental results is available at https://github.com/ericstrobl/RCINB.

 \subsection{Evaluation Criteria}
All of the above algorithms output an expected Shapley value for each patient and each variable. Moreover, the Shapley values involve a predictive model fit on the error terms. We therefore compared the outputs of the algorithms utilizing the root mean squared error (RMSE): 
$$\sqrt{\frac{1}{qp} \sum_{j=1}^q \sum_{i=1}^p (\widehat{\Upsilon}_{i}^j - \Upsilon_i^j)^2},$$ 
where lower is better. If an algorithm only estimates expected Shapley values for a subset of variables, then we set the values of the missing variables to zero. We computed the ground truth Shapley values $\Upsilon^j$ to negligible error by running XGBoost TreeSHAP on 100,000 ground truth values of $(\bm{E}, \bm{P}, D_0)$. We also measured the average running time of each algorithm in seconds.

\subsection{Synthetic Data}
\subsubsection{Data Generation}
We generated structural equation models obeying Equation \eqref{eq:PNL} as follows. We first generated DAGs with $p=7$ or $p=12$ variables in $\bm{X}$ and an expected neighborhood size of two. We created a random adjacency matrix by sampling from a Bernoulli$(2/(p-1))$ distribution in the upper triangular portion of the matrix. We introduced weights $\beta$ and offsets $\gamma$ by drawing values from the uniform distribution on $[-1,-0.25]$ in order to stay within machine precision after exponentiation -- except for the terminal vertex $D$ whose incoming edges $\beta_D$ were drawn from the uniform distribution on $[-1,-0.25] \cup [0.25,1]$. We drew $D$ randomly from the set of terminal vertices with at least one parent. We similarly set the shape parameters of each gamma distribution by drawing from a uniform distribution on $[0.1, 1]$. We drew $C$ from a gamma distribution with shape and rate equal to one. We generated $n=10,000$ or $100,000$ cell samples from $2$ to $10$ individuals in $\bm{P}$ again sampled uniformly. We repeated the above procedure 250 times and therefore generated a total of $250 \times 2 \times 2 = 1000$ independent datasets.

\subsubsection{Results}
We summarize the results in Table 1. Bolded values denote the best performance per dimension and sample size. All bolded values are significant at a Bonferonni corrected threshold of 0.05/5 using paired t-tests, since we compared the performance of five algorithms.

RCI-NB achieved the lowest mean RMSE across all dimension numbers and sample sizes. Moreover, the algorithm continued to improve with increasing sample sizes. The other algorithms all performed worse but comparably across dimensions; their performances also did not improve with increasing sample sizes. Accounting for Poisson measurement error thus steadily improves performance with more samples.

RCI and ICA both completed within two seconds due to the computational efficiencies gained by assuming linearity. RCI-NB took significantly longer than both RCI and ICA but completed approximately two times as quickly as the alternative non-linear approaches ANM and HNM. We conclude that RCI-NB is slower than linear methods but faster than alternative non-linear ones.

\subsection{Real Data}

\subsubsection{Lung Adenocarcinoma}
We evaluated the algorithms on their ability to discover the root causes of lung adenocarcinoma. The GSE-123904 scRNA-seq dataset of \cite{Laughney20} contains RNA counts from 17,502 single cells derived from cancerous and normal adjacent tissue of three patients (IDs 675, 682 and 684). The mitogen-activated protein kinase (MAPK) pathway plays an important role in lung carcinogenesis \cite{Kanehisa20}. KRAS and EGFR comprise the top driver genes in lung adenocarcinoma \cite{Martinez20,Liu20}. EGFR had nearly all zero counts in the data, so we included downstream genes GRB2, HRAS, ARAF and CCND1 instead. We added KRAS and TP53 as well.

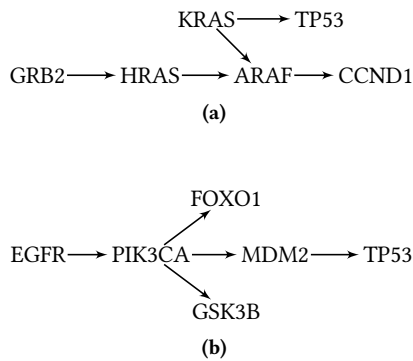
\begin{figure}[t]
\centering

\begin{subfigure}{0.49\textwidth}
\centering
\begin{tikzpicture}[scale=1.0, shorten >=1pt,auto,node distance=2.8cm, semithick]
                    
\tikzset{vertex/.style = {inner sep=0.4pt}}
\tikzset{edge/.style = {->,> = latex'}}
 
\node[vertex] (1) at  (0,0) {GRB2};
\node[vertex] (2) at  (1.5,0) {HRAS};
\node[vertex] (3) at  (3,0) {ARAF};
\node[vertex] (4) at  (4.5,0) {CCND1};
\node[vertex] (5) at  (2.25,0.75) {KRAS};
\node[vertex] (6) at  (3.75,0.75) {TP53};

\draw[edge] (1) to (2);
\draw[edge] (2) to (3);
\draw[edge] (3) to (4);
\draw[edge] (5) to (3);
\draw[edge] (5) to (6);

\end{tikzpicture}
\caption{} \label{fig_lung}
\end{subfigure}

\vspace{8mm}
\begin{subfigure}{0.49\textwidth}
\centering
\begin{tikzpicture}[scale=1.0, shorten >=1pt,auto,node distance=2.8cm, semithick]
                    
\tikzset{vertex/.style = {inner sep=0.4pt}}
\tikzset{edge/.style = {->,> = latex'}}
 
\node[vertex] (1) at  (0,0) {EGFR};
\node[vertex] (2) at  (1.5,0) {PIK3CA};
\node[vertex] (3) at  (3.15,0) {MDM2};
\node[vertex] (4) at  (4.65,0) {TP53};
\node[vertex] (5) at  (2.5,0.75) {FOXO1};
\node[vertex] (6) at  (2.5,-0.75) {GSK3B};

\draw[edge] (1) to (2);
\draw[edge] (2) to (3);
\draw[edge] (3) to (4);
\draw[edge] (2) to (5);
\draw[edge] (2) to (6);

\end{tikzpicture}
\caption{} \label{fig_cerv}
\end{subfigure} 
\caption{Ground truth causal graphs for (a) lung adenocarcinoma and (b) HPV16+ cervical squamous cell carcinoma.}
\end{figure}

We do not have access to the ground truth expected Shapley values with real data. However, we can estimate them to high accuracy using the ground truth causal graph. We obtained the causal relations between the genes using the KEGG pathway of non-small cell lung cancer (HSA05223) and plot the pathway in Figure \ref{fig_lung} \cite{Kanehisa00}. We estimated the ground truth Shapley values by (1) fitting negative binomial regression models using the sample versions of Equation \eqref{eq:asymp} on the ground truth parent set of each variable, (2) sampling from the de-noised gamma distributions and (3) running XGBoost TreeSHAP on the data of each patient. Recall that RCI-NB, ANM and HNM all use NB-EM, LR-EM and TreeSHAP.

\begin{figure}[t]
\begin{subfigure}[]{0.49\textwidth}
\captionsetup{justification=centering,margin=2cm}
    \centering
    \includegraphics[scale=0.45]{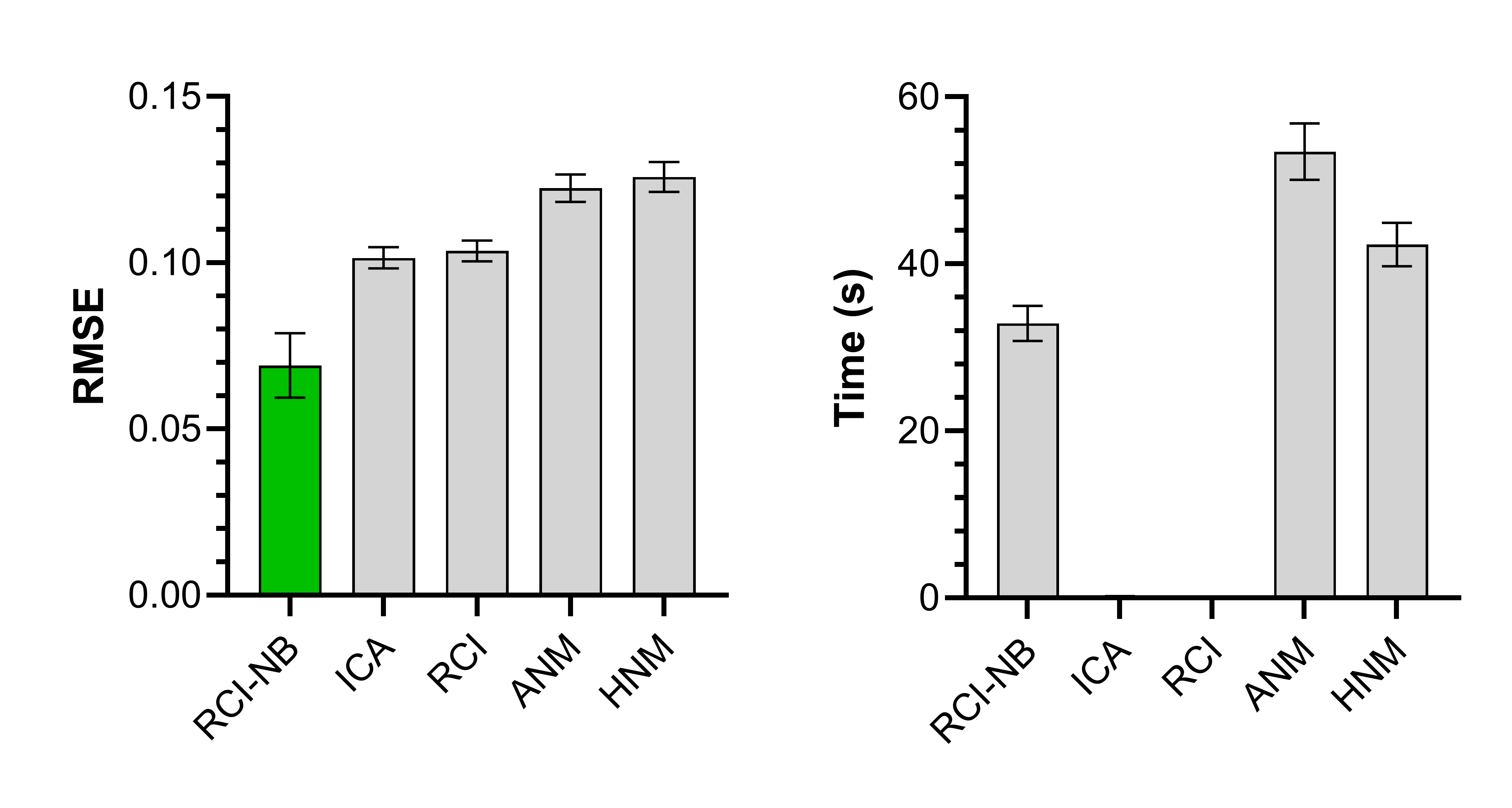}
    \caption{} \label{fig:lung}
\end{subfigure}
\begin{subfigure}[]{0.47\textwidth}
\captionsetup{justification=centering,margin=2cm}
    \centering
    \includegraphics[scale=0.45]{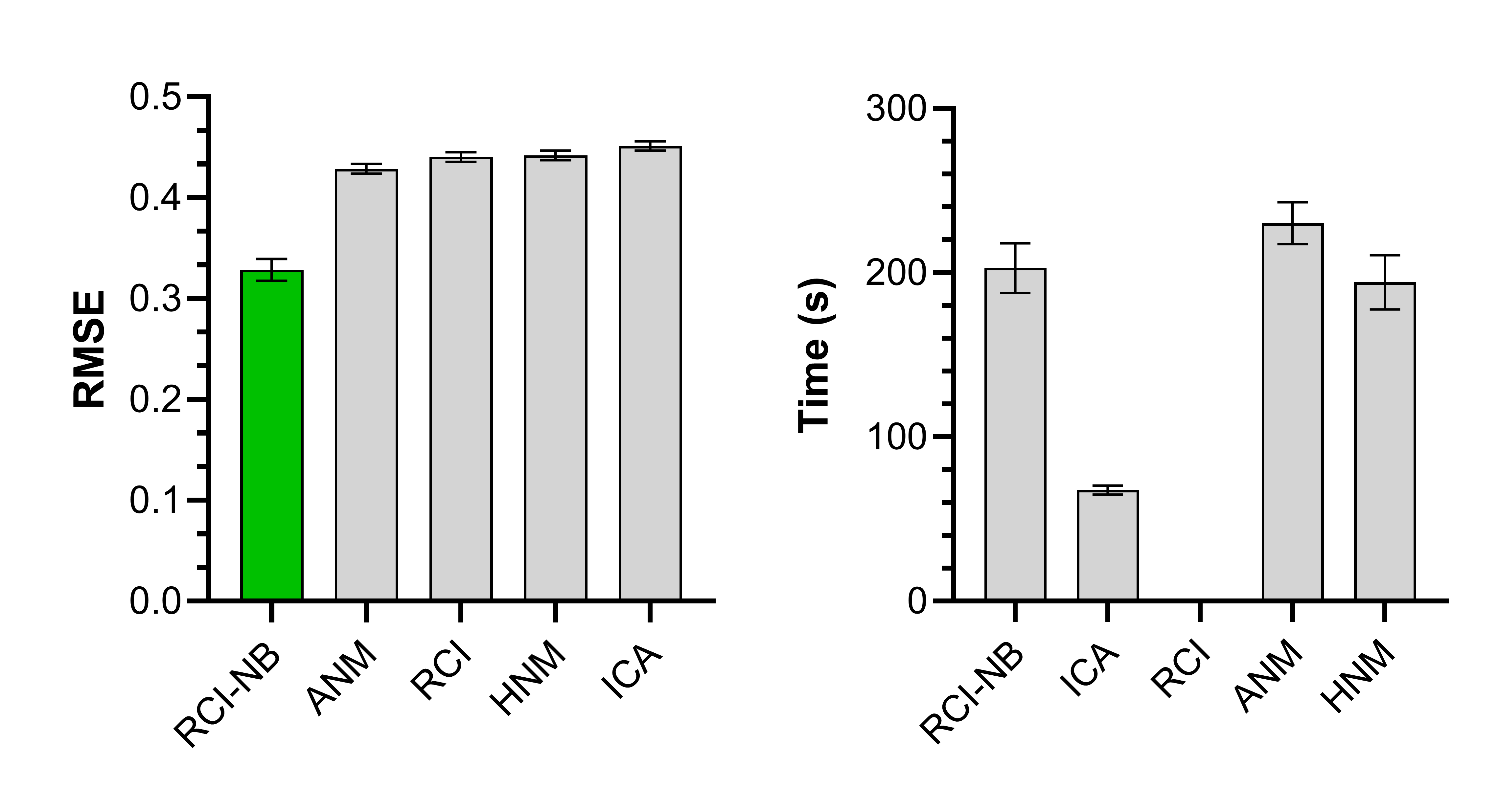}
    \caption{} \label{fig:cervical}
\end{subfigure}
\caption{Real data RMSE and timing results for scRNA-seq datasets derived from (a) lung adenocarcinoma and (b) HPV16+ cervical squamous cell carcinoma.}
\end{figure}

We plot the RMSE and timing results of the algorithms in Figure \ref{fig:lung} as averaged over 50 bootstrapped samples. RCI-NB achieved the lowest MSE by a large margin. The algorithms utilizing ANM and HNM achieved lower accuracy than the linear methods ICA and RCI. All non-linear algorithms took substantially longer than the linear ones, but RCI-NB still completed faster than both HNM and ANM. We conclude that RCI-NB achieves the highest accuracy and completes the fastest among the non-linear methods in this dataset.

\subsubsection{Cervical Carcinoma}

We next evaluated the ability of the algorithms to discover the root causes of cervical squamous cell carcinoma. We downloaded scRNA-seq data from E-MTAB-11948 used in \cite{Li22}. The dataset contains 69,938 cells from cancerous and normal adjacent tissue of three patients with cervical cancer. PIK3CA is the most frequently mutated gene in cervical carcinoma \cite{Martinez20,Liu20}. All three patients also tested positive for HPV type 16 that produces oncoproteins E5, E6 and E7 known to effect EGFR and the PI3K signaling pathway \cite{Li22,Hemmat20}. The PI3K signaling pathway effects cell cycle progression via GSK3B and FOXO1 as well as cell survival via MDM2 and TP53 according to the HPV KEGG pathway (HSA05165). We plot the ground truth causal graph in Figure \ref{fig_cerv}.

We summarize the results in Figure \ref{fig:cervical} as averaged over 50 bootstrapped draws. RCI-NB again achieved the lowest average RMSE by a large margin. The linear algorithms did not consistently outperform non-linear ANM and HNM. Instead, all algorithms besides RCI-NB performed comparably. RCI-NB took 202.7 seconds to complete on average, on-par with ANM and HNM in this case. We conclude that RCI-NB again achieves the highest accuracy with timing comparable to other non-linear algorithms. The real data results therefore mimic those seen with synthetic data.

\section{Conclusion}
We presented a post non-linear SEM consisting of gamma distributed error terms and random variables corrupted by Poisson measurement error. We then showed that each variable admits a negative binomial distribution when conditioned on its parents and patient. We used this fact to derive novel regression and goodness of fit testing procedures that bypass Poisson measurement error. The test requires samples from the joint distribution of the parents, which we recovered using the top-down RCI-NB algorithm. Experimental results highlighted the superiority of RCI-NB in recovering the true root causal contributions -- quantified using expected Shapley values -- in both synthetic and real data. Future work could improve the scalability of the method and accommodate latent confounding not related to measurement error.

\bibliography{biblio}


\begin{thebibliography}{38}


\ifx \showCODEN    \undefined \def \showCODEN     #1{\unskip}     \fi
\ifx \showDOI      \undefined \def \showDOI       #1{#1}\fi
\ifx \showISBNx    \undefined \def \showISBNx     #1{\unskip}     \fi
\ifx \showISBNxiii \undefined \def \showISBNxiii  #1{\unskip}     \fi
\ifx \showISSN     \undefined \def \showISSN      #1{\unskip}     \fi
\ifx \showLCCN     \undefined \def \showLCCN      #1{\unskip}     \fi
\ifx \shownote     \undefined \def \shownote      #1{#1}          \fi
\ifx \showarticletitle \undefined \def \showarticletitle #1{#1}   \fi
\ifx \showURL      \undefined \def \showURL       {\relax}        \fi
\providecommand\bibfield[2]{#2}
\providecommand\bibinfo[2]{#2}
\providecommand\natexlab[1]{#1}
\providecommand\showeprint[2][]{arXiv:#2}

\bibitem[Andersen and Fagerhaug(2006)]%
        {Anderson06}
\bibfield{author}{\bibinfo{person}{Bj{\o}rn Andersen} {and}
  \bibinfo{person}{Tom Fagerhaug}.} \bibinfo{year}{2006}\natexlab{}.
\newblock \bibinfo{booktitle}{\emph{Root cause analysis: simplified tools and
  techniques}}.
\newblock \bibinfo{publisher}{Quality Press}, \bibinfo{address}{United States}.
\newblock


\bibitem[Arzalluz-Luque et~al\mbox{.}(2017)]%
        {Arzalluz17}
\bibfield{author}{\bibinfo{person}{{\'A}ngeles Arzalluz-Luque},
  \bibinfo{person}{Guillaume Devailly}, \bibinfo{person}{Anna Mantsoki}, {and}
  \bibinfo{person}{Anagha Joshi}.} \bibinfo{year}{2017}\natexlab{}.
\newblock \showarticletitle{Delineating biological and technical variance in
  single cell expression data}.
\newblock \bibinfo{journal}{\emph{The International Journal of Biochemistry \&
  Cell Biology}}  \bibinfo{volume}{90} (\bibinfo{year}{2017}),
  \bibinfo{pages}{161--166}.
\newblock


\bibitem[Boos(1992)]%
        {Boos92}
\bibfield{author}{\bibinfo{person}{Dennis~D Boos}.}
  \bibinfo{year}{1992}\natexlab{}.
\newblock \showarticletitle{On generalized score tests}.
\newblock \bibinfo{journal}{\emph{The American Statistician}}
  \bibinfo{volume}{46}, \bibinfo{number}{4} (\bibinfo{year}{1992}),
  \bibinfo{pages}{327--333}.
\newblock


\bibitem[Budhathoki et~al\mbox{.}(2021)]%
        {Budhathoki21}
\bibfield{author}{\bibinfo{person}{Kailash Budhathoki},
  \bibinfo{person}{Dominik Janzing}, \bibinfo{person}{Patrick Bloebaum}, {and}
  \bibinfo{person}{Hoiyi Ng}.} \bibinfo{year}{2021}\natexlab{}.
\newblock \showarticletitle{Why did the distribution change?}. In
  \bibinfo{booktitle}{\emph{International Conference on Artificial Intelligence
  and Statistics}}. PMLR, \bibinfo{pages}{1666--1674}.
\newblock


\bibitem[Budhathoki et~al\mbox{.}(2022)]%
        {Budhathoki22}
\bibfield{author}{\bibinfo{person}{Kailash Budhathoki}, \bibinfo{person}{Lenon
  Minorics}, \bibinfo{person}{Patrick Bl{\"o}baum}, {and}
  \bibinfo{person}{Dominik Janzing}.} \bibinfo{year}{2022}\natexlab{}.
\newblock \showarticletitle{Causal structure-based root cause analysis of
  outliers}. In \bibinfo{booktitle}{\emph{International Conference on Machine
  Learning}}. PMLR, \bibinfo{pages}{2357--2369}.
\newblock


\bibitem[Choudhary and Satija(2022)]%
        {Choudhary22}
\bibfield{author}{\bibinfo{person}{Saket Choudhary} {and}
  \bibinfo{person}{Rahul Satija}.} \bibinfo{year}{2022}\natexlab{}.
\newblock \showarticletitle{Comparison and evaluation of statistical error
  models for scRNA-seq}.
\newblock \bibinfo{journal}{\emph{Genome Biology}} \bibinfo{volume}{23},
  \bibinfo{number}{1} (\bibinfo{year}{2022}), \bibinfo{pages}{27}.
\newblock


\bibitem[Ferguson(2017)]%
        {Ferguson17}
\bibfield{author}{\bibinfo{person}{Thomas~S Ferguson}.}
  \bibinfo{year}{2017}\natexlab{}.
\newblock \bibinfo{booktitle}{\emph{A course in large sample theory}}.
\newblock \bibinfo{publisher}{Routledge}, \bibinfo{address}{Boca Raton}.
\newblock


\bibitem[Guo et~al\mbox{.}(2001)]%
        {Guo01}
\bibfield{author}{\bibinfo{person}{Jie~Q Guo}, \bibinfo{person}{Tong Li},
  {et~al\mbox{.}}} \bibinfo{year}{2001}\natexlab{}.
\newblock \showarticletitle{Simulation-based estimation of the structural
  errors-in-variables negative binomial regression model with an application}.
\newblock \bibinfo{journal}{\emph{Annals of Economics and Finance}}
  \bibinfo{volume}{2}, \bibinfo{number}{1} (\bibinfo{year}{2001}),
  \bibinfo{pages}{101--122}.
\newblock


\bibitem[Hafemeister and Satija(2019)]%
        {Hafemeister19}
\bibfield{author}{\bibinfo{person}{Christoph Hafemeister} {and}
  \bibinfo{person}{Rahul Satija}.} \bibinfo{year}{2019}\natexlab{}.
\newblock \showarticletitle{Normalization and variance stabilization of
  single-cell RNA-seq data using regularized negative binomial regression}.
\newblock \bibinfo{journal}{\emph{Genome Biology}} \bibinfo{volume}{20},
  \bibinfo{number}{1} (\bibinfo{year}{2019}), \bibinfo{pages}{296}.
\newblock


\bibitem[He et~al\mbox{.}(2021)]%
        {He21}
\bibfield{author}{\bibinfo{person}{Liang He}, \bibinfo{person}{Jose
  Davila-Velderrain}, \bibinfo{person}{Tomokazu~S Sumida},
  \bibinfo{person}{David~A Hafler}, \bibinfo{person}{Manolis Kellis}, {and}
  \bibinfo{person}{Alexander~M Kulminski}.} \bibinfo{year}{2021}\natexlab{}.
\newblock \showarticletitle{NEBULA is a fast negative binomial mixed model for
  differential or co-expression analysis of large-scale multi-subject
  single-cell data}.
\newblock \bibinfo{journal}{\emph{Communications Biology}} \bibinfo{volume}{4},
  \bibinfo{number}{1} (\bibinfo{year}{2021}), \bibinfo{pages}{629}.
\newblock


\bibitem[Hemmat et~al\mbox{.}(2020)]%
        {Hemmat20}
\bibfield{author}{\bibinfo{person}{Nima Hemmat}, \bibinfo{person}{Ahad
  Mokhtarzadeh}, \bibinfo{person}{Mohammad Aghazadeh}, \bibinfo{person}{Farhad
  Jadidi-Niaragh}, \bibinfo{person}{Behzad Baradaran}, {and}
  \bibinfo{person}{Hossein Bannazadeh~Baghi}.} \bibinfo{year}{2020}\natexlab{}.
\newblock \showarticletitle{Role of microRNAs in epidermal growth factor
  receptor signaling pathway in cervical cancer}.
\newblock \bibinfo{journal}{\emph{Molecular Biology Reports}}
  \bibinfo{volume}{47} (\bibinfo{year}{2020}), \bibinfo{pages}{4553--4568}.
\newblock


\bibitem[Hwang et~al\mbox{.}(2018)]%
        {Hwang18}
\bibfield{author}{\bibinfo{person}{Byungjin Hwang}, \bibinfo{person}{Ji~Hyun
  Lee}, {and} \bibinfo{person}{Duhee Bang}.} \bibinfo{year}{2018}\natexlab{}.
\newblock \showarticletitle{Single-cell RNA sequencing technologies and
  bioinformatics pipelines}.
\newblock \bibinfo{journal}{\emph{Experimental \& Molecular Medicine}}
  \bibinfo{volume}{50}, \bibinfo{number}{8} (\bibinfo{year}{2018}),
  \bibinfo{pages}{1--14}.
\newblock


\bibitem[Kanehisa and Goto(2000a)]%
        {Kanehisa20}
\bibfield{author}{\bibinfo{person}{Minoru Kanehisa} {and}
  \bibinfo{person}{Susumu Goto}.} \bibinfo{year}{2000}\natexlab{a}.
\newblock \showarticletitle{KEGG: kyoto encyclopedia of genes and genomes}.
\newblock \bibinfo{journal}{\emph{Nucleic Acids Research}}
  \bibinfo{volume}{28}, \bibinfo{number}{1} (\bibinfo{year}{2000}),
  \bibinfo{pages}{27--30}.
\newblock


\bibitem[Kanehisa and Goto(2000b)]%
        {Kanehisa00}
\bibfield{author}{\bibinfo{person}{Minoru Kanehisa} {and}
  \bibinfo{person}{Susumu Goto}.} \bibinfo{year}{2000}\natexlab{b}.
\newblock \showarticletitle{KEGG: kyoto encyclopedia of genes and genomes}.
\newblock \bibinfo{journal}{\emph{Nucleic Acids Research}}
  \bibinfo{volume}{28}, \bibinfo{number}{1} (\bibinfo{year}{2000}),
  \bibinfo{pages}{27--30}.
\newblock


\bibitem[Lasko and Mesa(2019)]%
        {Lasko19}
\bibfield{author}{\bibinfo{person}{Thomas~A Lasko} {and}
  \bibinfo{person}{Diego~A Mesa}.} \bibinfo{year}{2019}\natexlab{}.
\newblock \showarticletitle{Computational Phenotype Discovery via Probabilistic
  Independence}.
\newblock \bibinfo{journal}{\emph{KDD Workshop on Applied Data Science for
  Healthcare}} (\bibinfo{year}{2019}).
\newblock


\bibitem[Laughney et~al\mbox{.}(2020)]%
        {Laughney20}
\bibfield{author}{\bibinfo{person}{Ashley~M Laughney}, \bibinfo{person}{Jing
  Hu}, \bibinfo{person}{Nathaniel~R Campbell}, \bibinfo{person}{Samuel~F
  Bakhoum}, \bibinfo{person}{Manu Setty}, \bibinfo{person}{Vincent-Philippe
  Lavallee}, \bibinfo{person}{Yubin Xie}, \bibinfo{person}{Ignas Masilionis},
  \bibinfo{person}{Ambrose~J Carr}, \bibinfo{person}{Sanjay Kottapalli},
  {et~al\mbox{.}}} \bibinfo{year}{2020}\natexlab{}.
\newblock \showarticletitle{Regenerative lineages and immune-mediated pruning
  in lung cancer metastasis}.
\newblock \bibinfo{journal}{\emph{Nature Medicine}} \bibinfo{volume}{26},
  \bibinfo{number}{2} (\bibinfo{year}{2020}), \bibinfo{pages}{259--269}.
\newblock


\bibitem[Li et~al\mbox{.}(2022)]%
        {Li22}
\bibfield{author}{\bibinfo{person}{Chunbo Li}, \bibinfo{person}{Hao Wu},
  \bibinfo{person}{Luopei Guo}, \bibinfo{person}{Danyang Liu},
  \bibinfo{person}{Shimin Yang}, \bibinfo{person}{Shengli Li}, {and}
  \bibinfo{person}{Keqin Hua}.} \bibinfo{year}{2022}\natexlab{}.
\newblock \showarticletitle{Single-cell transcriptomics reveals cellular
  heterogeneity and molecular stratification of cervical cancer}.
\newblock \bibinfo{journal}{\emph{Communications Biology}} \bibinfo{volume}{5},
  \bibinfo{number}{1} (\bibinfo{year}{2022}), \bibinfo{pages}{1208}.
\newblock


\bibitem[Liu et~al\mbox{.}(2020)]%
        {Liu20}
\bibfield{author}{\bibinfo{person}{Shu-Hsuan Liu}, \bibinfo{person}{Pei-Chun
  Shen}, \bibinfo{person}{Chen-Yang Chen}, \bibinfo{person}{An-Ni Hsu},
  \bibinfo{person}{Yi-Chun Cho}, \bibinfo{person}{Yo-Liang Lai},
  \bibinfo{person}{Fang-Hsin Chen}, \bibinfo{person}{Chia-Yang Li},
  \bibinfo{person}{Shu-Chi Wang}, \bibinfo{person}{Ming Chen}, {et~al\mbox{.}}}
  \bibinfo{year}{2020}\natexlab{}.
\newblock \showarticletitle{DriverDBv3: a multi-omics database for cancer
  driver gene research}.
\newblock \bibinfo{journal}{\emph{Nucleic Acids Research}}
  \bibinfo{volume}{48}, \bibinfo{number}{D1} (\bibinfo{year}{2020}),
  \bibinfo{pages}{D863--D870}.
\newblock


\bibitem[Liu and Li(2016)]%
        {Liu16}
\bibfield{author}{\bibinfo{person}{Zhenqiu Liu} {and} \bibinfo{person}{Gang
  Li}.} \bibinfo{year}{2016}\natexlab{}.
\newblock \showarticletitle{Efficient regularized regression with penalty for
  variable selection and network construction}.
\newblock \bibinfo{journal}{\emph{Computational and Mathematical Methods in
  Medicine}}  \bibinfo{volume}{2016} (\bibinfo{year}{2016}).
\newblock


\bibitem[Liu et~al\mbox{.}(2017)]%
        {Liu17}
\bibfield{author}{\bibinfo{person}{Zhenqiu Liu}, \bibinfo{person}{Fengzhu Sun},
  {and} \bibinfo{person}{Dermot~P McGovern}.} \bibinfo{year}{2017}\natexlab{}.
\newblock \showarticletitle{Sparse generalized linear model with L 0
  approximation for feature selection and prediction with big omics data}.
\newblock \bibinfo{journal}{\emph{BioData Mining}} \bibinfo{volume}{10},
  \bibinfo{number}{1} (\bibinfo{year}{2017}), \bibinfo{pages}{1--12}.
\newblock


\bibitem[Lundberg et~al\mbox{.}(2018)]%
        {Lundberg18}
\bibfield{author}{\bibinfo{person}{Scott~M Lundberg},
  \bibinfo{person}{Gabriel~G Erion}, {and} \bibinfo{person}{Su-In Lee}.}
  \bibinfo{year}{2018}\natexlab{}.
\newblock \showarticletitle{Consistent individualized feature attribution for
  tree ensembles}.
\newblock \bibinfo{journal}{\emph{arXiv preprint arXiv:1802.03888}}
  (\bibinfo{year}{2018}).
\newblock


\bibitem[Lundberg and Lee(2017)]%
        {Lundberg17}
\bibfield{author}{\bibinfo{person}{Scott~M Lundberg} {and}
  \bibinfo{person}{Su-In Lee}.} \bibinfo{year}{2017}\natexlab{}.
\newblock \showarticletitle{A unified approach to interpreting model
  predictions}. In \bibinfo{booktitle}{\emph{Proceedings of the 31st
  International Conference on Neural Information Processing Systems}}.
  \bibinfo{pages}{4768--4777}.
\newblock


\bibitem[Mart{\'\i}nez-Jim{\'e}nez et~al\mbox{.}(2020)]%
        {Martinez20}
\bibfield{author}{\bibinfo{person}{Francisco Mart{\'\i}nez-Jim{\'e}nez},
  \bibinfo{person}{Ferran Mui{\~n}os}, \bibinfo{person}{In{\'e}s Sent{\'\i}s},
  \bibinfo{person}{Jordi Deu-Pons}, \bibinfo{person}{Iker Reyes-Salazar},
  \bibinfo{person}{Claudia Arnedo-Pac}, \bibinfo{person}{Loris Mularoni},
  \bibinfo{person}{Oriol Pich}, \bibinfo{person}{Jose Bonet},
  \bibinfo{person}{Hanna Kranas}, {et~al\mbox{.}}}
  \bibinfo{year}{2020}\natexlab{}.
\newblock \showarticletitle{A compendium of mutational cancer driver genes}.
\newblock \bibinfo{journal}{\emph{Nature Reviews Cancer}} \bibinfo{volume}{20},
  \bibinfo{number}{10} (\bibinfo{year}{2020}), \bibinfo{pages}{555--572}.
\newblock


\bibitem[Nakamura(1990)]%
        {Nakamura90}
\bibfield{author}{\bibinfo{person}{Tsuyoshi Nakamura}.}
  \bibinfo{year}{1990}\natexlab{}.
\newblock \showarticletitle{Corrected score function for errors-in-variables
  models: Methodology and application to generalized linear models}.
\newblock \bibinfo{journal}{\emph{Biometrika}} \bibinfo{volume}{77},
  \bibinfo{number}{1} (\bibinfo{year}{1990}), \bibinfo{pages}{127--137}.
\newblock


\bibitem[Nelder and Wedderburn(1972)]%
        {Nelder72}
\bibfield{author}{\bibinfo{person}{John~Ashworth Nelder} {and}
  \bibinfo{person}{Robert~WM Wedderburn}.} \bibinfo{year}{1972}\natexlab{}.
\newblock \showarticletitle{Generalized linear models}.
\newblock \bibinfo{journal}{\emph{Journal of the Royal Statistical Society:
  Series A (General)}} \bibinfo{volume}{135}, \bibinfo{number}{3}
  (\bibinfo{year}{1972}), \bibinfo{pages}{370--384}.
\newblock


\bibitem[Newey and McFadden(1994)]%
        {Newey94}
\bibfield{author}{\bibinfo{person}{Whitney~K Newey} {and}
  \bibinfo{person}{Daniel McFadden}.} \bibinfo{year}{1994}\natexlab{}.
\newblock \showarticletitle{Large sample estimation and hypothesis testing}.
\newblock \bibinfo{journal}{\emph{Handbook of Econometrics}}
  \bibinfo{volume}{4} (\bibinfo{year}{1994}), \bibinfo{pages}{2111--2245}.
\newblock


\bibitem[Papoulis and Unnikrishna~Pillai(2002)]%
        {Papoulis02}
\bibfield{author}{\bibinfo{person}{Athanasios Papoulis} {and}
  \bibinfo{person}{S Unnikrishna~Pillai}.} \bibinfo{year}{2002}\natexlab{}.
\newblock \bibinfo{booktitle}{\emph{Probability, Random Variables and
  Stochastic Processes}}.
\newblock


\bibitem[Park and Raskutti(2017)]%
        {Park17}
\bibfield{author}{\bibinfo{person}{Gunwoong Park} {and}
  \bibinfo{person}{Garvesh Raskutti}.} \bibinfo{year}{2017}\natexlab{}.
\newblock \showarticletitle{Learning Quadratic Variance Function (QVF) DAG
  Models via OverDispersion Scoring (ODS)}.
\newblock \bibinfo{journal}{\emph{Journal of Machine Learning Research}}
  \bibinfo{volume}{18} (\bibinfo{year}{2017}), \bibinfo{pages}{224--1}.
\newblock


\bibitem[Rayner et~al\mbox{.}(2009)]%
        {Rayner09}
\bibfield{author}{\bibinfo{person}{John~CW Rayner}, \bibinfo{person}{Olivier
  Thas}, {and} \bibinfo{person}{Donald~John Best}.}
  \bibinfo{year}{2009}\natexlab{}.
\newblock \bibinfo{booktitle}{\emph{Smooth tests of goodness of fit: using R}}.
\newblock \bibinfo{publisher}{John Wiley and Sons}, \bibinfo{address}{Germany}.
\newblock


\bibitem[Sarkar and Stephens(2021)]%
        {Sarkar21}
\bibfield{author}{\bibinfo{person}{Abhishek Sarkar} {and}
  \bibinfo{person}{Matthew Stephens}.} \bibinfo{year}{2021}\natexlab{}.
\newblock \showarticletitle{Separating measurement and expression models
  clarifies confusion in single-cell RNA sequencing analysis}.
\newblock \bibinfo{journal}{\emph{Nature genetics}} \bibinfo{volume}{53},
  \bibinfo{number}{6} (\bibinfo{year}{2021}), \bibinfo{pages}{770--777}.
\newblock


\bibitem[Schwarz(1978)]%
        {Schwarz78}
\bibfield{author}{\bibinfo{person}{Gideon Schwarz}.}
  \bibinfo{year}{1978}\natexlab{}.
\newblock \showarticletitle{Estimating the dimension of a model}.
\newblock \bibinfo{journal}{\emph{Annals of Statistics}}
  (\bibinfo{year}{1978}), \bibinfo{pages}{461--464}.
\newblock


\bibitem[Strobl and Lasko(2022a)]%
        {Strobl22a}
\bibfield{author}{\bibinfo{person}{Eric~V. Strobl} {and}
  \bibinfo{person}{Thomas~A. Lasko}.} \bibinfo{year}{2022}\natexlab{a}.
\newblock \showarticletitle{Identifying Patient-Specific Root Causes of
  Disease}. In \bibinfo{booktitle}{\emph{Proceedings of the 13th ACM
  International Conference on Bioinformatics, Computational Biology and Health
  Informatics}} (Northbrook, Illinois) \emph{(\bibinfo{series}{BCB '22})}.
  \bibinfo{publisher}{Association for Computing Machinery},
  \bibinfo{address}{New York, NY, USA}, Article \bibinfo{articleno}{18},
  \bibinfo{numpages}{10}~pages.
\newblock
\showISBNx{9781450393867}


\bibitem[Strobl and Lasko(2022b)]%
        {Strobl22b}
\bibfield{author}{\bibinfo{person}{Eric~V Strobl} {and}
  \bibinfo{person}{Thomas~A Lasko}.} \bibinfo{year}{2022}\natexlab{b}.
\newblock \showarticletitle{Identifying Patient-Specific Root Causes with the
  Heteroscedastic Noise Model}.
\newblock \bibinfo{journal}{\emph{arXiv preprint arXiv:2205.13085}}
  (\bibinfo{year}{2022}).
\newblock


\bibitem[Strobl and Lasko(2023)]%
        {Strobl23}
\bibfield{author}{\bibinfo{person}{Eric~V Strobl} {and}
  \bibinfo{person}{Thomas~A Lasko}.} \bibinfo{year}{2023}\natexlab{}.
\newblock \showarticletitle{Sample-Specific Root Causal Inference with Latent
  Variables}.
\newblock \bibinfo{journal}{\emph{Causal Learning and Reasoning}}
  (\bibinfo{year}{2023}).
\newblock


\bibitem[Svensson(2020)]%
        {Svensson20}
\bibfield{author}{\bibinfo{person}{Valentine Svensson}.}
  \bibinfo{year}{2020}\natexlab{}.
\newblock \showarticletitle{Droplet scRNA-seq is not zero-inflated}.
\newblock \bibinfo{journal}{\emph{Nature Biotechnology}} \bibinfo{volume}{38},
  \bibinfo{number}{2} (\bibinfo{year}{2020}), \bibinfo{pages}{147--150}.
\newblock


\bibitem[Wu et~al\mbox{.}(2008)]%
        {Wu08}
\bibfield{author}{\bibinfo{person}{Albert~W Wu}, \bibinfo{person}{Angela~KM
  Lipshutz}, {and} \bibinfo{person}{Peter~J Pronovost}.}
  \bibinfo{year}{2008}\natexlab{}.
\newblock \showarticletitle{Effectiveness and efficiency of root cause analysis
  in medicine}.
\newblock \bibinfo{journal}{\emph{Jama}} \bibinfo{volume}{299},
  \bibinfo{number}{6} (\bibinfo{year}{2008}), \bibinfo{pages}{685--687}.
\newblock


\bibitem[Zhang and Hyv{\"a}rinen(2009)]%
        {Zhang09}
\bibfield{author}{\bibinfo{person}{K Zhang} {and} \bibinfo{person}{A
  Hyv{\"a}rinen}.} \bibinfo{year}{2009}\natexlab{}.
\newblock \showarticletitle{On the Identifiability of the Post-Nonlinear Causal
  Model}. In \bibinfo{booktitle}{\emph{25th Conference on Uncertainty in
  Artificial Intelligence (UAI 2009)}}. AUAI Press, \bibinfo{pages}{647--655}.
\newblock


\bibitem[Ziegenhain et~al\mbox{.}(2022)]%
        {Ziegenhain22}
\bibfield{author}{\bibinfo{person}{Christoph Ziegenhain},
  \bibinfo{person}{Gert-Jan Hendriks}, \bibinfo{person}{Michael
  Hagemann-Jensen}, {and} \bibinfo{person}{Rickard Sandberg}.}
  \bibinfo{year}{2022}\natexlab{}.
\newblock \showarticletitle{Molecular spikes: a gold standard for single-cell
  RNA counting}.
\newblock \bibinfo{journal}{\emph{Nature Methods}} \bibinfo{volume}{19},
  \bibinfo{number}{5} (\bibinfo{year}{2022}), \bibinfo{pages}{560--566}.
\newblock


\end{thebibliography}

\section*{Appendix}
\subsection{Proposition 1}
Let $\theta=(\alpha,r)$ and $X_i = Y$. Consider the derivative of the corrected log-likelihood given by $\frac{1}{n} \sum_{i=1}^n S_i(\theta)$ where $\theta = (\alpha, r)$ and:
\begin{equation} \nonumber
\begin{aligned}
S_i(\alpha) &= y_i \bm{z}_i - \frac{1}{s}\sum_{j=1}^s \mu_j \widetilde{\bm{z}}_j\\
S_i(r) &= \psi(y_i + r) - \psi(r) + \textnormal{ln}(r) - \frac{1}{s} \sum_{j=1}^s \textnormal{ln}(r + \mu_j).
\end{aligned}
\end{equation}
The original uncorrected versions of the score equations correspond to:
\begin{equation} \nonumber
\begin{aligned}
S_i^*(\alpha) &= y_i \bm{z}_i - \frac{r + y_i}{r + \mu_i} \mu_i \bm{z}_i\\
S_i^*(r) &= 1 -\frac{r + y_i}{r + \mu_i} + \psi(y_i + r) - \psi(r) + \textnormal{ln}(r) - \textnormal{ln}(r + \mu_i),
\end{aligned}
\end{equation}
The following conclusion holds:
\begin{repproposition1}{prop:normality}
(Asymptotic normality) Assume $n \rightarrow \infty, s \rightarrow \infty$ and $n/s \rightarrow 0$. Further assume that $\Omega = \textnormal{Var}(\mu \widetilde{\bm{Z}}, \textnormal{ln}(r+\mu))$ and $\Sigma = -\mathbb{E} S^\prime(\theta_0)$ are positive definite. Then $\sqrt{n}(\theta_n - \theta_0) \rightarrow \mathcal{N}(0,\Sigma^{-1}(J_1 + J_2 + J_3) \Sigma^{-1}).$
\end{repproposition1}
\begin{proof}
    We can write:
    \begin{equation} \label{eq:components}
        \frac{1}{\sqrt{n}} \sum_{i=1}^n S_i (\theta) = \frac{1}{\sqrt{n}} \sum_{i=1}^n S^*_i (\theta) + \frac{1}{\sqrt{n}} \sum_{i=1}^n A_i(\theta) + \frac{\sqrt{n}}{s} \sum_{j=1}^s B_j(\theta),
    \end{equation}
    where:
    \begin{equation} \nonumber
         A_i(\theta) =  \left( \begin{aligned}
            &\hspace{1.2cm}\frac{r + y_i}{r + \mu_i} \mu_i \bm{z}_i - \mathbb{E} \mu \widetilde{\bm{Z}}\\
            &\frac{r + y_i}{r + \mu_i} - 1 + \textnormal{ln}(r + \mu_i) - \mathbb{E} \textnormal{ln}(r + \mu)
        \end{aligned} \right),
    \end{equation}
    and:
    \begin{equation} \nonumber
         B_j(\theta) = \left(  \begin{aligned} &\hspace{7mm}\mathbb{E}\mu \widetilde{\bm{Z}} - \mu_j \widetilde{\bm{z}}_j\\ &\mathbb{E}\textnormal{ln}(r+\mu) - \textnormal{ln}(r+\mu_j)
        \end{aligned}\right).
    \end{equation}
    We consider a compact neighborhood $Q_\rho = \{ \theta : |\theta - \theta_0|  \leq \rho \}$ for some $\rho >0$. We now invoke the integral form of the mean valued theorem \cite[page 20]{Ferguson17}: $$\frac{1}{\sqrt{n}} \sum_{i=1}^n S_i (\theta_n) = \frac{1}{\sqrt{n}} \sum_{i=1}^n S_i (\theta_0) - \sqrt{n}(\theta_n - \theta_0)C_n,$$
    where $C_n = - \int_0^1 \frac{1}{n}\sum_{i=1}^n S^\prime_i (\theta_0 + u(\theta_n - \theta_0)) ~du$. 
    
     We have $\sup_{\theta \in Q_\rho} |\frac{1}{n} \sum_{i=1}^n S_i(\theta) - \mathbb{E}_{\theta_0}S(\theta)| \rightarrow 0$ almost surely by the uniform strong law of large numbers with $n \rightarrow \infty$ and $s \rightarrow \infty$ \cite[page 108]{Ferguson17}. We then invoke Theorem 2.1 in \cite[pages 2121-2122]{Newey94} to conclude that $\theta_n$ is a strongly consistent sequence satisfying $\sum_{i=1}^n S_i (\theta_n) = 0$. Therefore $\frac{1}{\sqrt{n}} \sum_{i=1}^n S_i (\theta_0) = \sqrt{n}(\theta_n - \theta_0)C_n$.
    
    We consider the right hand side of Equation \eqref{eq:components}. We have: $\frac{1}{\sqrt{n}} \sum_{i=1}^n S^*_i (\theta_0) \leadsto \mathcal{N}(0,J_1),$ where $J_1 = \mathbb{E} S^*(\theta_0) S^{*T}(\theta_0)$. We also have $ \frac{1}{\sqrt{n}} \sum_{i=1}^n A_i (\theta_0) \leadsto \mathcal{N}(0,J_2),$
    where $J_2 = \mathbb{E} A(\theta_0) A^T(\theta_0).$ Thus: $\frac{1}{\sqrt{n}} \sum_{i=1}^n S^*_i (\theta_0) + A_i (\theta_0) \leadsto \mathcal{N}(0,J_1+J_2 + J_3),$
    where $J_3 =  \mathbb{E} A(\theta_0)S^{*T}(\theta_0) + \mathbb{E} S^{*}(\theta_0)A^T(\theta_0).$ We finally have: $\frac{1}{\sqrt{s}} \sum_{j=1}^s B_j (\theta_0) \leadsto \mathcal{N}(0,\Omega),$ since $\Omega$ is positive definite. We invoke Slutsky's lemma so that: $\frac{\sqrt{n}}{s} \sum_{j=1}^s B_j (\theta_0) = \sqrt{\frac{n}{s}}\frac{1}{\sqrt{s}} \sum_{j=1}^s B_j (\theta_0) \leadsto 0,$
    because $n/s \rightarrow 0$. As a result: $\frac{1}{\sqrt{n}} \sum_{i=1}^n S_i (\theta_0) \leadsto \mathcal{N}(0,J_1 + J_2 + J_3).$ 
    
    We next show that $C_n \rightarrow \Sigma$ almost surely. We let $\varepsilon > 0$. The function $\mathbb{E}_{\theta_0} S^\prime(\theta) $ is continuous in $\theta$. We can therefore identify a $\rho >0$ such that $| \theta - \theta_0 | < \rho$ implies $| \mathbb{E}_{\theta_0} S^\prime (\theta) + \Sigma | < \varepsilon/2$.

    Again by the uniform strong law of large numbers, there exists an integer $N$ such that the following holds with probability one for all $n > N: \sup_{\theta \in Q_\rho} \Big|  \frac{1}{n} \sum_{i=1}^n S_i^\prime(\theta) - \mathbb{E}_{\theta_0} S^\prime(\theta) \Big| < \varepsilon/2$. Now assume $N$ is so large such that for all $n > N$, we have $| \theta_n - \theta_0 | < \rho$. Hence, for all $n > N$, we have:
    \begin{equation} \nonumber
    \begin{aligned}
        &| C_n - \Sigma |  \leq \int_0^1 \Big| \frac{1}{n} \sum_{i=1}^n S^\prime_i (\theta_0 + u(\theta_n - \theta_0)) + \Sigma \Big|~du\\
        \leq \hspace{1mm}& \int_0^1  \sup_{\theta \in Q_\rho} \Big|  \frac{1}{n} \sum_{i=1}^n S_i^\prime(\theta) - \mathbb{E}_{\theta_0} S^\prime(\theta) \Big| + | \mathbb{E}_{\theta_0} S^\prime(\theta) + \Sigma | ~du < \varepsilon.
    \end{aligned}
    \end{equation}
    We conclude that $C_n \rightarrow \Sigma$ almost surely because we chose $\varepsilon$ arbitrarily. We now invoke Slutsky's lemma:
    \begin{equation} \nonumber
        \sqrt{n}(\theta_n - \theta_0) = C_n^{-1} \frac{1}{\sqrt{n}} \sum_{i=1}^n S_i(\theta_0) \leadsto \mathcal{N}(0,\Sigma^{-1}(J_1 + J_2 + J_3) \Sigma^{-1}).
    \end{equation}
\end{proof}

\subsection{Theorem 1}
We consider the following overdispersion score:
$$T(X_i, \widetilde{\bm{U}}) = w^2_{i\widetilde{\bm{U}}} \textnormal{Var}(X_i | \widetilde{\bm{U}}) - w_{i \widetilde{\bm{U}}} \mathbb{E}(X_i | \widetilde{\bm{U}}),$$
where $w^{-1}_{i\widetilde{\bm{U}}} = 1 + \mathbb{E}(X_i | \widetilde{\bm{U}})/r$ and $\widetilde{\bm{U}} \subseteq \widetilde{\bm{X}} \cup \bm{P} \cup C$. The negative binomial of $X_i$ conditional on $\widetilde{\bm{U}}$ has mean $\mathbb{E}(X_i | \widetilde{\bm{U}})$ and variance $\mathbb{E}(X_i | \widetilde{\bm{U}}) + \mathbb{E}(X_i | \widetilde{\bm{U}})^2/r$. Thus $T(X_i, \widetilde{\bm{U}}) = 0$ in this case. 

Let $\widetilde{\bm{U}}$ more specifically correspond to a subset of the non-descendants of $X_i$ always including $\bm{P} \cup C$. We have:
\begin{lemma1} \label{lem:S}
If $\mathbb{P}_{\widetilde{\bm{X}}|\bm{P}}$ is causally minimal and $X_i \sim \textnormal{Pois}(\widetilde{X}_i C)$ for each $X_i \in \bm{X}$, then $T(X_i, \widetilde{\bm{U}}) = 0$ if and only if $\textnormal{Pa}(\widetilde{X}_i)\subseteq \widetilde{\bm{U}}$.
\end{lemma1}
\begin{proof}
We can write the following sequence:
\begin{equation} \nonumber
\begin{aligned}
&T(X_i, \widetilde{\bm{U}}) = w^2_{i\widetilde{\bm{U}}} \textnormal{Var}(X_i | \widetilde{\bm{U}}) - w_{i \widetilde{\bm{U}}} \mathbb{E}(X_i | \widetilde{\bm{U}})\\
&\stackrel{(a)}{=}w^2_{i\widetilde{\bm{U}}} \Big[   
 \textnormal{Var}(\mathbb{E}(X_i | \textnormal{Pa}(\widetilde{X}_i),\bm{P},C)|\widetilde{\bm{U}}) + \mathbb{E}(\textnormal{Var}(X_i|\textnormal{Pa}(\widetilde{X}_i),\bm{P},C)|\widetilde{\bm{U}})\\ &\hspace{6cm} - w^{-1}_{i \widetilde{\bm{U}}} \mathbb{E}(X_i|\widetilde{\bm{U}})\Big]\\
&\stackrel{(b)}{=}w^2_{i\widetilde{\bm{U}}} \Big[   
 \textnormal{Var}(\mathbb{E}(X_i | \textnormal{Pa}(\widetilde{X}_i),\bm{P},C)|\widetilde{\bm{U}}) + \mathbb{E}(\mathbb{E}(X_i|\textnormal{Pa}(\widetilde{X}_i),\bm{P},C)|\widetilde{\bm{U}})\\&\hspace{1cm}+\mathbb{E}(\mathbb{E}(X_i|\textnormal{Pa}(\widetilde{X}_i),\bm{P},C)^2/r|\widetilde{\bm{U}}) - (1 + \mathbb{E}(X_i | \widetilde{\bm{U}})/r) \mathbb{E}(X_i|\widetilde{\bm{U}})\Big]\\
  &=w^2_{i\widetilde{\bm{U}}} \Big[   
 \textnormal{Var}(\mathbb{E}(X_i | \textnormal{Pa}(\widetilde{X}_i),\bm{P},C)|\widetilde{\bm{U}}) +\mathbb{E}(\mathbb{E}(X_i|\textnormal{Pa}(\widetilde{X}_i),\bm{P},C)^2/r|\widetilde{\bm{U}})\\ 
 &\hspace{6cm} - \mathbb{E}^2(X_i|\widetilde{\bm{U}})/r\Big] \\
 &=w^2_{i\widetilde{\bm{U}}} (1+1/r) \textnormal{Var}(\mathbb{E}(X_i | \textnormal{Pa}(\widetilde{X}_i),\bm{P},C)|\widetilde{\bm{U}}),
\end{aligned}
\end{equation}
where (a) follows from the variance decomposition formula, and (b) from the quadratic variance property of the negative binomial. 

For the backward direction, if $\textnormal{Pa}(\widetilde{X}_i)\subseteq \widetilde{\bm{U}}$, then\\ $\textnormal{Var}(\mathbb{E}(X_i | \textnormal{Pa}(\widetilde{X}_i),\bm{P},C)|\widetilde{\bm{U}}) = 0$, so $T(X_i, \widetilde{\bm{U}}) = 0$. For the forward direction, assume by contrapositive that $\widetilde{\bm{U}}$ does not contain all members of $\textnormal{Pa}(\widetilde{X}_i)$. Then $\textnormal{Var}(\mathbb{E}(X_i | \textnormal{Pa}(\widetilde{X}_i),\bm{P},C)|\widetilde{\bm{U}}) > 0$ by causal minimality, so  $T(X_i, \widetilde{\bm{U}}) > 0$.
\end{proof}

\begin{reptheorem1}{thm:identifiability}
If $\mathbb{P}_{\widetilde{\bm{X}}|\bm{P}}$ is causally minimal and $X_i \sim \textnormal{Pois}(\widetilde{X}_i C)$ for each $X_i \in \bm{X}$, then RP recovers $(\beta,\bm{r},\gamma)$ with regression and goodness of fit oracles.
\end{reptheorem1}
\begin{proof}
We prove the statement by induction. Base: suppose $|\bm{X}| = 1$. Then $X_i \in \bm{X}$ is a Poisson-gamma mixture and therefore a negative binomial. Hence, RP recovers $(\beta_{\cdot i},r_i,\gamma_{\cdot i})$ in Line \ref{alg_CDGoF:reg1}. 

Induction: suppose the conclusion holds with $|\widetilde{\bm{X}}| = p$. We need to prove the statement when $|\widetilde{\bm{X}}\cup \widetilde{X}_i| = p+1$. We have two situations:
\begin{enumerate}[leftmargin=*,label=(\arabic*)]
\item Assume that $\widetilde{\bm{A}}$ contains all of the parents of $\widetilde{X}_i$ and none of its descendants. Then $X_i$ is a Poisson-gamma mixture given $\widetilde{\bm{A}} \cup \bm{P} \cup C$ and hence a negative binomial. RP again recovers $(\beta_{\cdot i},r_i,\gamma_{\cdot i})$ in Line \ref{alg_CDGoF:reg1}. The algorithm then places $X_i$ into $\bm{A}$ and removes it from $\bm{X}$ in Line \ref{alg_CDGoF:change}. 
\item Assume that $\widetilde{\bm{A}}$ (a) does not contain all of the parents of $\widetilde{X}_i$ or (b) contains at least one of the descendants of $\widetilde{X}_i$ from $\widetilde{\bm{X}} \setminus \widetilde{X}_i$ (or both). For (a), assume for a contradiction that RP removes $X_i$ from $\bm{X}$ in Line \ref{alg_CDGoF:change}. Then $S(X_i, \widetilde{\bm{A}} \cup \bm{P} \cup C)>0$ by Lemma \ref{lem:S} but this contradicts the fact that $S(X_i, \widetilde{\bm{A}} \cup \bm{P} \cup C)=0$ because $X_i$ follows a negative binomial. Thus RP does not place $X_i$ into $\bm{A}$ and remove it from $\bm{X}$ in Line \ref{alg_CDGoF:change}. For (b), assume for a contradiction that $\widetilde{\bm{A}}$ contains a descendant of $\widetilde{X}_i$ from $\widetilde{\bm{X}} \setminus \widetilde{X}_i$. But then there exists at least one descendant $\widetilde{X}_j$ of $\widetilde{X}_i$ ($\widetilde{X}_j \not = \widetilde{X}_i$) whose parents are not all in $\widetilde{\bm{A}}$, and $\widetilde{X}_j$ was removed by RP in a previous iteration. We therefore arrive at a contradiction again by Lemma \ref{lem:S}. We conclude that RP does not place $X_i$ into $\bm{A}$ and remove it from $\bm{X}$ in Line \ref{alg_CDGoF:change} in either case. 
\end{enumerate}
The conclusion follows by the inductive hypothesis.
\end{proof}

\end{document}